\documentclass{article}
\pdfoutput=1
\usepackage[utf8]{inputenc}
\usepackage{fullpage}
\usepackage{rotating}
\usepackage{graphicx}
\usepackage{multicol}
\usepackage{color}
\usepackage{multirow}
\usepackage[table]{xcolor}
\usepackage{longtable}
\usepackage{caption}
\usepackage{blindtext}
\usepackage{bm}
\usepackage{todonotes}
\usepackage{mathtools}
\usepackage{amsmath,amssymb}
\DeclareMathOperator{\E}{\mathbb{E}}
\usepackage{esint}
\usepackage{stmaryrd}
\usepackage{amsfonts}
\usepackage{dcolumn}
\usepackage{booktabs}
\usepackage{tablefootnote}
\usepackage{caption}
\usepackage{subcaption}
\usepackage{longtable}
\usepackage{hyperref}
\usepackage[authoryear, sort, numbers]{natbib}  
\setcitestyle{citesep={;}}
\usepackage{cleveref}
\usepackage{etoolbox}
\usepackage{algorithmic}
\usepackage{dirtytalk}
\usepackage{nicematrix} 
\usepackage[flushleft]{threeparttable}
\usepackage{threeparttablex}
\usepackage{centernot}
\usepackage[export]{adjustbox}
\usepackage{pdflscape}

\makeatletter

\patchcmd{\NAT@test}{\else \NAT@nm}{\else \NAT@nmfmt{\NAT@nm}}{}{}

\DeclareRobustCommand\citepos
  {\begingroup
   \let\NAT@nmfmt\NAT@posfmt
   \NAT@swafalse\let\NAT@ctype\z@\NAT@partrue
   \@ifstar{\NAT@fulltrue\NAT@citetp}{\NAT@fullfalse\NAT@citetp}}

\let\NAT@orig@nmfmt\NAT@nmfmt
\def\NAT@posfmt#1{\NAT@orig@nmfmt{#1's}}

\makeatother

\title{Peer Networks and Malleability of Educational Aspirations}
\author{Michelle González Amador\footnote{UNU-MERIT, Boschstraat 24, 6211AX, Maastricht, The Netherlands. E-mail: mgonzalez@merit.unu.edu}
\and
Robin Cowan\footnote{UNU-MERIT, Boschstraat 24, 6211AX, Maastricht, The Netherlands. E-mail: cowan@merit.unu.edu}
\and
Eleonora Nillesen\footnote{UNU-MERIT, Boschstraat 24, 6211AX, Maastricht, The Netherlands. E-mail: nillesen@merit.unu.edu}}
\date{\today}

\begin{document}

\maketitle
\begin{abstract}
    \noindent Continuing education beyond the compulsory years of schooling is one of the most important choices an adolescent has to make; higher education is associated with a host of social and economic benefits both for the person and its community. Today, there is ample evidence that educational aspirations are an important determinant of said choice. We implement a multilevel, networked experiment in 45 Mexican high schools, and provide evidence of the malleability of educational aspirations, and the interdependence of students' choices and the effect of our intervention with peer networks. Moreover, we find that a video-intervention, which combines role-models and information about returns to education, is successful in updating students' beliefs and consequently educational aspirations.  
\end{abstract} 
\vskip0.5em

\noindent\textbf{Keywords:} Aspirations, Education, Adolescents, Economics of Networks, Peer Effects, School transitions, Field Experiment, Social Network Analysis, Randomized Controlled Trial.
\vskip0.5em

\noindent\textbf{JEL Codes:} A21, C21, C22, C93, D83, D91, I29.

\section*{Acknowledgement}
This project would not have been possible without the administrative and financial support of the University of Guadalajara and the University of Colima, in Mexico. We thank Dr. Genoveva Amador Fierros, Director of International Relations, M. Ana Cecilia García Valencia, Coordinator of International Relations and Dr. Christian Jorge Torres Ortiz Zermeño, Rector of the University of Colima for their support in the state of Colima. We thank Dr. Carlos Iván Moreno Arellano, General Coordinator of the Academic and Innovation Department and M. Miguel Ángel Sigala Gómez, Coordinator of International Relations for their constant support in the city of Guadalajara, Jalisco. We acknowledge the continued efforts from both institutions to host our research, and the hard work from the Offices of International Relations and their staff. We also acknowledge the invaluable contributions from Dr. Francisco Marmolejo, president for Higher Education at the Qatar Foundation. His feedback and suggestions are present in the practical part of the design of this experiment. Heriberto Álvarez Núñez and Angélica Rosalía López Velazco, Economics and Finance students from the University of Guadalajara, who showed leadership and initiative throughout the data collection process.     

\section{Introduction}
\label{section:intro}
Continuing education beyond the compulsory years of schooling is perhaps one of the most important choices an adolescent has to make. Not only do we persistently find cross-country evidence of tertiary education providing young people with an economic premium in the labour market \citep{psacharopoulos2018returns}, but  higher education is also associated with a host of social benefits such as better health, good citizenship, and reduced crime rates \citep{lochner2011non}. Yet, high school to university transition rates remain low in many countries and students from disadvantaged populations are less likely to choose to go into higher education \citep{french2017behavioral}. Traditionally, research has pointed to policy investments in early-childhood educational interventions to increase transition rates and human capital \citep[e.g.][]{carneiro2003human}. However, recent studies are building the argument that policy interventions should target students during the transition period: adolescents’ decision-making differs from that of adults \citep{lavecchia2016behavioral,french2017behavioral}, and behavioural biases that are naturally exaggerated during this period can be addressed to encourage university enrolment \citep{koch2015behavioral, jabbar2011behavioral, damgaard2018nudging}. What is more, college-going interventions, in particular those that contain an element of student encouragement, have been found to have long-lasting effects on university enrollment and retainment \citep{carrell2017college}; the above is an indication that there is still a need for policy action at the high school level. 

\paragraph{}
In the spirit of understanding adolescents’ educational choices during transition periods, we combine three elements from their educational decision-making process, and design an intervention that aims to encourage human capital investment. First, we look at students’ educational aspirations as drivers of their intrinsic motivation to further their years of schooling, and focus on the malleability of aspirations to beget belief and behavioural change in enrollment choices.\footnote{Note that we only measure change in aspirations, and assume that this will later translate into enrollment choices} Students’ aspirations are commonly understood to be a proxy for goal-oriented behaviour with important implications for educational attainment. Recent studies have provided evidence of the malleability of aspirations by showing how exogenous shocks, such as institutional policy changes \citep{beaman2012female}, programme interventions \citep{garcia2019building} or relaxing financial constraints \citep{glewwe2018developing} affect aspirations so that children and adolescents not only aspire to more education, but also increase the number of years they study. Second, we look at students’ attention and expectations with regard to the benefits and costs of higher education. By now, it is largely accepted that information regarding educational benefits and opportunities does not reach  all members of society equally. Even with the move to store and handle information online, which has increased reach and transparency from universities and other formal institutions, students still rely on other, more direct sources to make up their minds about where and how long to study \citep[e.g.][]{akerlof2002identity,hoxby2013low}. That students still make their decisions with incorrect or partial information has not only prompted researchers to understand the sources of misinformation, but also to look for strategies to address informational gaps and observe the conditions under which they yield behavioural change. Finally, there is a long tradition of the study of peer effects in educational outcomes, from performance to enrollment \citep{sacerdote2011peer}. Recent evidence points to peers having a more important role in education than previously thought, affecting not only own and group average scores, but also the development of other relevant educational skills \citep{zarate2019social}. Moreover, there is evidence that behavioural interventions in a high school setting can be filtered through the school's social network in order to increase the probability of success, i.e. change general school stance and behavior regarding issues such as bullying \citep{paluck2016changing,shepherd2015stopping}.

\paragraph{}
We design a behavioural intervention that garners key characteristics from these three elements and implement it in a sample of 45 public high schools in the Midwestern region of Mexico. The schools vary in size, geographical location, and student composition, but form two large clusters, each  belonging to a different state in the region. They are Colima, known for its agricultural sector and trading port, and Guadalajara, colloquially referred to as the Silicon Valley of Latin America. These two locations differ in their main economic activity and labour market opportunities after schooling, which allows us to see whether the intervention functions differently under two different environments, and thus discuss the external validity of the tool. The intervention is a video-based story telling of success cases of former students from the region (i.e. role models) who successfully completed a university degree, alongside information about the private returns of different levels of education\footnote{We also compile a series of financial aid opportunities and leave leaflets in the schools’ main offices for students to collect for themselves if they so desire}. We inject the intervention into the schools’ social network, to observe both the change in the treated students and in the students' friendship network within the school, i.e. social and cognitive peer effects. Our aim is to contribute to the understanding of educational decision-making processes, especially with respect to students’ aspirations, knowledge about returns to education, and schools’ underlying peer dynamics. Our results  provide insights on informational gaps and interventions on schooling behavior and choices, helping us understand how behavioural interventions and technology (videos) can be leveraged to improve students’ motivation and/or change perceptions, and we show evidence of the validity of the theoretical base of our intervention across different environments.

\section{Literature Review}

Several recent studies demonstrate that cognitive biases, along with subjective measures of ability,  and (incomplete) information influence human capital investment choices of adolescents, and consequently, affect their future economic lives. \citet{jabbar2011behavioral}, \citet{lavecchia2016behavioral}, and \citet{koch2015behavioral} all document the different ways in which behavioural biases affect the educational decision-making of adolescents. In particular, \citet{lavecchia2016behavioral} dedicate a section to understanding how young people differ from adults, with a note on how the perceptions adolescents form about themselves, their abilities and identity, affect both the effort they make in school and the decisions of how long and where to study. The combination of these two student outputs has long-run effects on the accumulation of human and economic capital.  

\subsection{Malleability of Aspirations and its effect on educational outcomes and choices}
\label{subsection:malleabilityofasp}

There is a growing literature that is uncovering how self-perception, schooling attitudes, and the subsequent aspirations we form for our future directly affect our educational outcomes. Using longitudinal data for the United Kingdom, \citet{chowdry2011role} find that young people's attitudes in school, such as low aspirations and expectations for future education, and low self-confidence correlate with low educational attainment. They also note that a high incidence of these negative behaviours can be found with students from a low socioeconomic background. The findings from \citet{chowdry2011role} are in line with earlier work in psychology, where it is found that self-efficacy, or the beliefs about our own abilities, is both a determinant of our educational outcomes and a driver of our aspirations \citep[e.g.][]{bandura1977self,bandura2003negative}, with work in economics on the unequal distribution of aspirations by socioeconomic strata \citep[e.g.][]{ray2006aspirations,guyon2021biased,dalton2016poverty} and with work on students' imperfect information about own ability \citep{benabou2003intrinsic}. In the same vein, \citep{wydick2013does}, carry out a cross-country analysis of the effect of a socio-emotional skill development programme with a strong focus on self-esteem and aspirations, and find that an increase in these skills leads to an increase in secondary school completion. Similar observational work that relates aspirations, expectations and other forms of socio-emotional skills on academic performance (and choices) can be found in reviews by \citet{heckman2006effects}, \citet{goodman2010poorer}, and \citet{gorard2012impact}.   

\paragraph{}
In economics, experimental studies shed light on the malleable nature of aspirations. In particular, the work by \citet{beaman2012female}, \citet{garcia2019building}, and \citet{riley2017increasing} stand out in the field of education. \citet{beaman2012female} exploit a natural experiment in the West Bengal region of India, where a gender quota policy was gradually introduced for city councils. They find that in the villages where women were elected into the local councils, young girls not only increased their educational and labor market aspirations, but as a consequence also increased their educational attainment, as compared to the young girls in villages where there was no quota and no women reached the city council. They argue that, by exposing young girls to real-life role models, they were able to visualize a different future for themselves and increase their aspirations. \citet{garcia2019building} also expose adolescents, and their parents, to role models. They invite beneficiaries of the Colombian cash transfer, Familias en Acción, to meetings with local leaders and career professionals. In these meetings, they discuss the benefits of education and longer schooling on labour market outcomes. They find that the social interaction between role models and beneficiares increases both aspirations of young people and that of their parents. Moreover, they say that the observed increase in educational attainment of the children who benefit from the cash transfer program can be partially explained by higher aspirations. Relatedly, \citet{riley2017increasing} design an experiment in Uganda, where a sample of secondary school children were exposed to a role model by watching a motivational film, while a control group watched a placebo movie. The Disney film, Queen of Katwe, tells the story of a young girl from the Ugandan slum of Katwe who works hard to learn the game of chess. She slowly starts to win against players from all over the world and all walks of life. Riley's analysis shows that treated students increased their academic performance in a mathematics exam. Furthermore, the study emphasizes the the role of technological innovations (e.g. films) to decrease the costs associated with socio-emotional interventions in educational settings. 

\paragraph{}
These studies highlight the relevance of measuring subjective beliefs on ability, aspirations, and other forms of motivation and self-perception when trying to unravel the mechanisms of human capital investment. This information provides a basis for the design of educational interventions that aim to increase academic engagement, academic performance \citep[e.g.][]{riley2017increasing,nguyen2008information} and influence schooling choice \citep[e.g.][]{beaman2012female}. They show us that engaging young people with inspiring, relatable and diverse thoughts, in the form of role models, can help them aspire to a better educational future, and mobilise their cognitive resources to achieve this goal. Furthermore, they show that, in conjunction with simple technology, we can create this effect in a cost-effective way \citep{riley2017increasing,mani2019social}.       

\subsection{Attention, expectations and belief formation in regards to returns to education}

Educational aspirations and expectations are not formed in a vacuum. Students exist in a school-community cosmos where information about the state of the(ir) world comes from different sources, and they learn and arrive at aspirational beliefs from processing this information. In this regard, \citet{damgaard2018nudging} and \citet{french2017behavioral} pioneer a new discussion about the implications of a world with asymmetric information and young people's limited attention and cognitive space when choosing how and when to educate themselves. Even when information is publicly available, knowledge on the benefits of schooling, including public and private returns to education, is not correctly shared across subgroups of adolescents or their families \citep{scott2015role,french2017behavioral}. Evidence on gendered \citep{attanasio2014education}, racial \citep{akerlof2002identity}, and socioeconomic \citep{jensen2010perceived, kaufmann2014understanding} clustering of (mis)information regarding returns to education begs the question of how adolescent decision-makers interact with a complex system of information sharing. 

\paragraph{}
\citet{damgaard2018nudging} propose that limited attention and complex social and educational systems make students susceptible to biased and/or suboptimal decision-making. That is, they overlook information around them and make choices based on partial knowledge. For example, \citet{dynarski2006cost} and \citet{bettinger2012role} show that complexity in bureaucracy discourages students from seeking economic aid for educational purposes, even when they express interest in doing so. The same is true regarding the verification of information about schooling opportunities, schooling costs, and returns to schooling. In the United States, \citet{hoxby2013low} find that high-achieving students from a low socioeconomic background fail to apply to selective universities, despite the high probability of acceptance and lower tuition costs, because they follow the advice and trends from people in their community rather than acquiring the information from the universities themselves. Reviewing interventions that aim to reduce information asymmetries on returns to schooling, \citet{french2017behavioral} document how sharing correct information on costs and benefits \citep[e.g.][with a video-intervention]{dinkelman2014investing} changes students' attitudes towards enrollment to tertiary education, albeit with small effect sizes. In Mexico, 
\citet{attanasio2014education}, show that subjective expectations about returns to education are particularly important for predicting tertiary enrollment, and especially so for boys. Moreover, using Chilean data, \citet{hastings2016informed} show that while overestimating initial costs to higher education discourages enrollment and increases the probability of dropping out when students do enroll, overestimating returns to education increases the probability of choosing a low-return major and of defaulting on loans. In a Low and Middle Income (LMIC) country setting, the probability of having incorrect information about returns to education is high for students with a low socioeconomic background, and their perceived monetary return is an important predictor of enrollment \citep{attanasio2014education,jensen2010perceived}. 
However, in their review \citet{french2017behavioral} find no clear link between reducing the information gap and actual enrollment behavior. For instance, whilst \citet{oreopoulos2013information} and \citet{dinkelman2014investing} find a positive effect of sharing post-secondary schooling benefits and financial aid information on enrollment choices of high school students in Canada and Chile respectively, both \citet{hastings2016informed} and \citet{busso2017effects} find no effect of providing said information on college enrollment rates of Chilean high school students. 
A possible explanation for the mixed results on the effects of information provision on students' outcomes could lie in the aspirations literature. That is, there exist complementarities between students' educational aspirations and what they do or how they process the information they are given.   
\citet{nguyen2008information} follows a similar logic as she looks at test scores (effort in school) of students. She uses an experimental setup where she presents a group of students with a video containing information on returns to education, another similar sample of students is presented with a video with role models, and a third sample watches a video combining both tools. She finds that both the role model and the statistics videos have a positive effect on test scores (and other measures of effort, such as school attendance), but that the treatment effect of the combination of both was smaller than for the video with only statistics (on returns to education). Nguyen's study demonstrates that both role models and videos are effective tools to raise students' effort in school; however, due to the controlled environment embedded in her experimental design, she cannot draw conclusions about the effect of her combined treatment on students' outcomes in non-controlled or dynamic setups. Students experience social interactions with peers on a daily basis. Could these alter the experienced effects in Nguyen's sample? There is a need to explore how the delivery of information and exposure to role models function under the context of social dynamics. 

\subsection{Peer effects, social identities and peer networks}  

As students develop an understanding of the dynamics of the social and academic environments they navigate, peer interactions become a convenient source of information and a reinforcement mechanism during the process of belief formation. It has been extensively documented that peers influence social and academic behavior of adolescents, ranging from risky and criminal behavior such as alcohol consumption or bullying, to academic achievement, including test scores and enrollment choices \citep[for a review, see][]{sacerdote2011peer}. Whilst the vast majority of research on peer effects in education is concerned with the identification of a causal effect between the individual and the average outcome of the group, the growing popularity of experiments \citep{sacerdote2011peer} and the introduction of social network methodologies \citep{angrist2014perils} has helped research transition from identification of causality to the exploration of the mechanisms at play.

\paragraph{}
The emerging framework in peer effects in education is one that is concerned with how students integrate the input of their immediate society (peers) into their decision-making, and how the characteristics of peers and peer groups shape social interactions and educational outcomes. The first idea refers to how students form social identities and make choices based on social image concerns. In this vein, \citet{fryer2006acting} proposes that a sense of belonging to a (racial) group imposes a psychological cost on deviating from the behavior of the group, with negative consequences for educational investments for, say, black people in the United States who have historically had fewer opportunities regarding education programmes. A person would have to get a sizeable (pecuniary) gain from misbehaving - investing in education and exiting their community- to offset the loss of their social ties and ethnic identity. \citet{benabou2011identity} discuss how people invest in their identity, and the actions attached to the phenotype of their identity, as a way to both reconcile morality and social integration. Experimentally, \citet{bursztyn2015does} and \citet{bursztyn2019cool} find that schooling investments are discouraged when students' actions are observable by peers. \citet{bursztyn2019cool} show that the mechanism behind this is students' concerns about social image. They describe how in a school where smart students are lauded and their actions are observable by peers, under-performing students who are afraid to be seen as non-smart are discouraged from taking up free preparatory courses for University. Taking up the course puts them at risk of under-performing and having their non-achievement be observed by the rest of the student population. Similarly, in a school where \say{cool} behaviour (disinterest in academic achievement) is praised by students and actions are observable by peers, academically inclined students are discouraged from taking up university preparatory courses because they do not want to be seen as \say{trying too hard} in academic endeavours.

\paragraph{}
Furthermore, to understand how peers' characteristics and group composition affect schooling outcomes, \citet{dieye2014accounting} refer to \citepos*{bramoulle2009identification} introduction of a social network analysis for peer effect identification. In particular, \citet{dieye2014accounting} mention that by correctly identifying actors and connections without restricting interactions outside of an observable cluster (e.g. classroom), we can differentiate between endogenous and contextual peer effects, i.e. effects that correspond to peers' outcomes as a function of the group's outcome, and effects that correspond to peers' outcomes given the proportion of friends who experience a shock (or treatment), respectively. \citepos{dieye2014accounting} paper is a good example as to why this differentiation matters.
They analyse the effect of a government issued cash transfer on children's school attendance. First, they demonstrate that the treatment effect on the treated is positive, i.e. children whose families receive a cash transfer are more likely to attend school. They also find that peer attendance is (statistically significant and small but) positive; that is, endogenous peer effects are positive. They say that this is consistent with theories on `positive complementarities between peers in time spent in social activities outside school' \citep{dieye2014accounting}. However, as the proportion of treated friends increases, own school attendance decreases; that is, they find negative contextual peer effects. They propose that his finding is due to the fact that student cash transfer beneficiaries and their peers are substitutes in the child labour market. In other words, the contextual peer effects are driven by students who are not beneficiaries, and who are aware of the freed up labour opportunities that their treated peers gave up. Their findings show that, not only is there no benefit from a social multiplier effect of a randomly assigned programme as documented in the literature \citep[e.g.][]{glaeser2003social,bobonis2009neighborhood}\footnote{We explicitly say no benefit, because the positive results from the endogenous peer effect indicates that there exists a social multiplier.}, but the net effect of the programme is close to zero when accounting for the negative contextual peer effects.

\paragraph{}
In an experiment, \citet{feld2017understanding} look at the effect of peer group composition on students' academic achievement when students are randomly assigned to sections at the university level. They find suggestive evidence that the positive, non-linear (endogenous) peer effect on grades is the result of improved group interaction. In a follow up study, \citet{golsteyn2021impact} show that the suggested improved group interaction is due to the presence of more persistent and more risk-averse peers. \citet{zarate2019social} reintroduces social network analysis to focus on peers' characteristics in more detail. He uses the random allocation of students to student dorms in a Peruvian boarding high school to study the development of personality traits and its effect on grades. He finds that in dorms with more sociable peers (identified through their position in the friendship network), students develop better social skills, the effect being driven by self-confidence; however, he finds no effect of social skills on academic achievement. Finally, \citet{shepherd2015stopping} and \citet{paluck2016changing} randomly select a sample of highly connected peers in high schools and invite them to participate in an anti-conflict intervention. The aim of the intervention was to beget behavioural change with regard to school harassment and bullying. \citet{shepherd2015stopping} find that \say{popular}\footnote{Popular students were identified as nodes in the 90th percentile of strength and closure scores in the school's network.} students in a high school peer network, for whom the treatment changed individual stance on the topic, helped reduce negative behaviours and harassment throughout the school thanks to their relative social influence. \citet{paluck2016changing} scale up the intervention to 56 high schools and observe a decrease of conflict of 30 per cent in treated schools, effectively capitalising on popular peers' capacity to leverage their social ties for community behavioral change. 
 
\paragraph{}  
The studies referenced above highlight the relevance of peer networks and their the capacity to influence the success or failure of schooling programmes. Social interactions, in a school setting, induce individual behavior that in some ways goes against actions the student may act upon in isolation. Thus, manipulating peer dynamics might offer an opportunity for positive or desirable behavior in schools and, in turn, help in the implementation of a programme that aims to better students' outcomes. 

\section{Methodology}
\subsection{Experimental Design}
\label{subsection:ExperimentalDesign}
We conduct a multilevel, networked experiment to determine the effect of an aspirational video-intervention and peers' inputs on the malleability of educational aspirations. 
We introduce a hierarchical design where a set of students, denoted by $A^s$, is semi-randomly selected to watch an aspirational video. Let $A^s = \{1,\ldots,k\}$, where $k \leq N^s/4$. $A^s$ is nested in 1 of 45 public high schools $s$. Therefore, $s = \{1,\ldots,45\}$, and $N^s$ denotes the total student population $N$ in school $s$.\footnote{The proportions used in the experiment can be found in the Appendix, \cref{subsection:School-attributes} \cref{tab:school-prop}.} Additionally, schools are matched to create comparable pairs. One school in the matched pair is randomly assigned to a treatment group, and its counterpart to a control group, such that $z^s \in \{0,1\}$ indicates whether school $s$ belongs to the treatment group or not.The final matched treatment assignment can be found in \cref{subsection:School-attributes} \cref{tab:school-match}. We use pre-treatment information for the matching, including school size, target cohort size, and the school-level clustering coefficient.

\paragraph{}
The aim of this design is twofold. First, the semi-aleatory selection of students $A^s$ allows us to optimise the identification of influential peers in school $s$' social network. We do so through sample stratification by gender and betweenness centrality, and defining a selection cutoff of above-median strength \citep[a strategy similar to that of][]{paluck2016changing}. As a note, `betweenness' measures the centrality of a student in the overall communication network, and `strength' is a weighted measure of the number of friends a student has. Second, the only students that receive the treatment are those in $A^s$ \& $z^s$ simultaneously; that is, only students eligible for treatment in treated schools. Therefore, we can estimate treatment spillovers through the students' social network in $A^s$, and draw insights on the role of peers on the effect of the video-intervention. We estimate individual-level treatment effects by comparing treated students to treatment-eligible students from non-treated schools. We estimate treatment spillovers by comparing students in treated schools to students in controls school, irrespective of students' treatment status. We expand on the empirical estimation in \cref{subsubsection: Causality_and_peers}.  

\paragraph{Description of the Treatment:}
The treatment, otherwise referred to as `intervention', consists of a 15 minute video that combines factual information on the economic returns to educational degrees in Mexico and an aspirational message.\footnote{The video was an ad-hoc development from footage captured by local Colima videographer Alma Méndez, and edited by Alma Méndez and René González Chávez. The file is available upon request.} The aspirational component of the video is a storytelling of three resilience cases, as experienced by former students from neighbouring communities \citep[echoing other aspirational video-interventions, such as][]{lybbert2018poverty,bernard2014future}. The stories showcase the struggles the `role-models' had during their transition from high school to university; including financial setbacks, health issues and family losses. Importantly, they discuss how they framed their obstacles as challenges to overcome\footnote{These cases are meant to convey the relevance of setting goals and aspiring to higher education in order to achieve significant changes in your life, regardless of the circumstances you are born into or the obstacles that may arise along the way.}. The informational component of the video discusses the pecuniary benefits attached to different levels of educational attainment, from high school (the minimum threshold for the target audience) to postgraduate school, including Master and Ph.D. levels. It is important to note that the data in the video, obtained from the National Institute of Statistics and Geography (INEGI), show decreasing returns to education. Students reach their peak salary after a Master's degree, and earn slightly less if they continue for a Ph.D. The purpose of combining these two components in the video is to update beliefs about the economic returns to schooling and help students make informed decisions about their educational careers.

\paragraph{Development of the treatment:}

To understand how a video-intervention with a motivational component affects a target audience, it is customary to have a control group that is presented with a placebo video.The placebo video must be comparable in format and convey similar emotions, if not necessarily the exact message. Should students in the control group respond in the same way as those in the treatment group, one would not be able to claim that the contents in the video-intervention matter beyond the screening of positive emotions\citep{bernard2015will}.  \citet{bernard2015will} explain that, in the context of LMICs where information asymmetries are larger, a placebo film allows researchers to assess the extent to which the video-information they share is relevant beyond being exposed to positive emotions with a scarce technology. \citet{bernard2014future} implement an aspirtional video-intervention in the rural Ethiopian context, following \citet{bernard2015will}'s placebo film approach. They find that both treated and placebo people respond positively in a host of outcomes of interest, but the magnitude of the effect is larger for treated individuals than for placebo individuals. \citet{banerjee2019entertainment} uses a similar placebo film approach in the context of a video-intervention to decrease the acceptance of gender based violence. They find significant effects on information and a decrease in gender violence acceptance, but no similar trajectory in the control group. \citet{banerjee2019entertaining} use the same educational video-intervention as as \citet{banerjee2019entertainment} and the same placebo video, but in the context of information on HIV/AIDS and changing uninformed attitudes and risky sexual behaviour. Their placebo, the `Gidi up' series, is a Nigerian web drama series.  The relevance of video-placebos in the context of video-interventions in LMICs can be seen as a way to disentangle the effect of the content of the video from the use of the technology - which may be a source of excitement for some (sub)populations, i.e. a primary concern in the literature, and thus argument for a placebo design, is fear that simply exposure to a novel technology affects the responses; and as a way to establish similar conditions or environments between experimental conditions. In the midwestern Mexican context, the latter is not an issue (nor possible confound), as video technology is neither scarce nor novel. However, by choosing not to have a placebo film we run the risk that the changes in behaviour we note for the treated students arise not from the content of the  film but simply from the positive emotions they feel from being excused from class and watching a video. There is no a priori reason to believe a change in emotional state will have a strong effect on our outcome of interest. While placebo films are indeed the standard \citep{bernard2015will}, the field of psychology provides ample evidence that changes in emotional states induced by subtle cues (read: a change in emotional state induced by out of classroom time and a placebo film; i.e. social priming) are unlikely to to have a large influence on behaviour \citep[for a review, see:][]{schimmack2021invalid}, and especially so in between-subject experimental designs such as the present one \citep{rivers2018experimental}. 

\paragraph{}
We deviate from the placebo approach, and establish a control group that is not exposed to targeted video-information but that is informed about the existence of informational leaflets in their school administration building.\footnote{The primary drawback of a non-placebo control is that we are unable to assess the extent to which the magnitude of our treatment effect(s) is due to the content of the video and how much of it is due to the elicitation of positive emotions from ``watching a film during class time''. On the other hand, we benefit from comparing a group that receives targeted information and a group that mimics real life, where adolescents are expected to seek the information for themselves.}  We circumvent any issues that might derive from the effect being purely a consequence of eliciting positive emotions, by combining the aspirational component of the video with economic data. In addition, the data come with no message or instructions on which salary segment is more desirable. We emphasise that an important element in our design is its ability to highlight informational paths, both by comparing targeted information to habitual information-seeking behaviour, and by observing information flows through peers' social networks.
The video-intervention does preserve an important element highlighted by \citet{banerjee2019entertaining, banerjee2019entertainment} and \citet{bernard2015will}: the use of visual technology to capture subjects' attention and convey complex information in a relatable way. The use of digital resources in behavioural interventions is especially important when the target audience are adolescents \citep{yeager2019national}. The attention span of teenagers is especially short \citep{geri2017probing}, and digital technology makes it easier to keep interventions short and cost-effective \citep{riley2017increasing,yeager2019national}. 

\paragraph{}
\noindent Ultimately, we expose students to messages delivered by `role models' that beget vicarious emotions, and include clear information about the economic benefits of educational degrees\footnote{We use the word degree here to refer to an attained educational level. We do not include information on different fields of study.}. Our main focus is 
on understanding the interplay of the video-intervention and peer social networks. Additionally, we want to be able to compare networked behaviour in relation to information asymmetries, when information is targeted and non-targeted but available. Our video contains a motivational component and a pedagogical component that complement each other, and it is therefore not solely reliant on evoking emotions as more traditional video interventions \citep[as per][]{bernard2015will}. Similar to \citet{banerjee2019entertaining}, there is a larger theme (i.e. plot) that is reinforced with `role models' and shared information, rendering an engaging visual stimulus that relays more than a fleeting emotion. Thus, we measure treated and control students' sociability and analyze how it interacts with the video, i.e. behaviour diffusion.   

\paragraph{Description of the implementation:}
Students in their last year of high school, grade 12, are first introduced to the research project during the baseline data collection phase. Using a semi-aleatory strategy\footnote{Unlike traditional approaches, we don't target low-income students, but focus our targeting strategy on highly central students in the school network, regardless of their socioeconomic status.}, a group of students is selected from treatment schools to participate in a screening of a 15-minute video. These students, who represent up to 25 per cent of their cohort, are then directly surveyed. Their peers are surveyed thereafter, throughout the course of the same day.

\section{Data and Analysis}

\subsection{Data collection and processing}
We collect data from two sources, student surveys and administrative records. The observational unit is the student, i.e. students enrolled in grade 12, the transition year to university in Mexico, whose parents signed a permission slip and who also voluntarily agree to participate in the study. Enumerators were present to monitor that students had their permission slips.The data belongs to three different collection periods\footnote{Exact dates are not offered as not all schools were surveyed on the exact same day. However, they were surveyed in the same week.}:

\begin{itemize}
    \item[·] All students in grade 12 at $t = 0$ (October 2018): baseline survey and administrative data.
        \begin{itemize}
           \item[-]Enumerators start collecting   data at the beginning of the school   day and have until the end of the     school day to make sure all     (volunteering) students have been     surveyed. Each class is given 50 minutes in the Computer Classroom to respond to the survey. Professors coordinate with the enumerators and facilitate the process, by making sure that all classes have the opportunity to go to the Computer classroom as soon as it is vacated by a previous class. 
        \end{itemize}
    \item[·] All students in grade 12 at $t = 1$ (February/March 2019): survey and administrative data (After exposing selected students to the video).
        \begin{itemize}
            \item[-] Enumerators play the video-intervention for selected students at the beginning of the school day. Professors have been informed who the selected students are (treated students hereafter) a day prior. Professors and enumerators coordinate to bring them to a designated classroom for the screening. At the end of the screening, treated students are asked to respond to the midline survey in the Computer Classroom and then escorted back to their respective classrooms. At the end of the `intervention', each class is ushered to the Computer Classroom to complete the survey. Because treated students are \emph{not} kept from the rest of their classes during the day, some of them may have time to interact with non-treatment eligible students throughout the day and discuss the contents of the video. \textbf{Note} that this is not a problem in our experimental design for non-contaminated ATET estimates (cf. \cref{subsubsection:ExperimentalSubsample}), nor for the behavioural diffusion analysis.
        \end{itemize}
    \item[·] All students in grade 12 at $t = 2$ (May 2019): survey and administrative data. 
    \begin{itemize}
            \item[-]Enumerators start collecting   data at the beginning of the school day and have until the end of the school day to make sure all (volunteering) students have been surveyed. Each class is given 50 minutes in the Computer Classroom to respond to the survey. Professors coordinate with the enumerators and facilitate the process, by making sure that all classes have the opportunity to go to the Computer classroom as soon as it is vacated by a previous class\footnote{Unfortunately, many students were no longer present as the date was close to the summer holidays and some classes are no longer mandatory to pass a course. In \cref{subsection:attrition} we show this period is the main source of attrition. Fortunately, the attrition pattern is random and uncorrelated to the treatment status.}.  
        \end{itemize}
\end{itemize}
Due to the process in which treated and non-treated students were surveyed during the intervention period [as described above], some treated students may have had time to interact with non-treated peers throughout the day (but not all). The smallest surveyed school had two grade 12 classes, whereas the largest school had six. This means that, in the smallest school, half of the treated students would be able to interact with non-treated students for 35 minutes (the length of time it takes a class to respond to the survey); for the largest class, five sixths of the treated students would have at least 35 minutes and at most 175 minutes (re: 1h55min). While we cannot guarantee students will not (mischievously) chat during class time, we can think of the intensity of this interaction as somewhat contained. Our data does not have classroom cluster identifiers, so we are unable to say anything about the difference between classroom behaviour by interaction length.However, should peer sociability throughout the day be important, we should be able to notice a difference in change in aspirations among students between the smallest and the largest school. That is, there should be more students changing their aspirations in the largest school, where there is more time between the screening and the last surveyed classroom, granting students more opportunity for mischievous chatter, than in the smallest school. We find that this is not the case, with about 45\% of students changing their aspirations at time [01] in the smallest school (Bachillerato 14, with 22 students), and 37\% of students changing their aspirations at time [01] in the largest school (Bachillerato 2, with 316 students. 

\paragraph{\textbf{Survey data:}}
The baseline survey took place in October 2018, with the sampled students being surveyed again at midline in February/March 2019 and endline in May 2019, throughout to the duration of 2018/2019 school year. The survey was carried out by research assistants that joined the study from undergraduate programs in Economics, Psychology and Social Work from the local universities, University of Guadalajara and University of Colima. 
 
The principals of each school designated a teacher who accompanied the research assistants and helped mobilize students to the schools’ computer centres, where they were introduced to the study by the research assistant and with an online consent form before starting the online survey. The baseline survey was first written in English and then translated to Spanish. It had a duration of 45 minutes, and was administered in the schools’ computer centres via a Qualtrics online link. The midline and  followup surveys both had an estimated answer time of 25 minutes.

\paragraph{\textbf{Administrative data:}}
We retrieved data on students' grades from the local universities' records. Data on grades correspond to the final students' grades for semester 5 and the final grades for semester 6 (corresponding to the two semesters that compose grade 12), and are obtained under a confidentiality agreement, in partnership with the Research Department from the University of Colima and the University of Guadalajara. 

\subsection{Description of the data}

We sample students from the Midwestern Region of Mexico, in the adjoining states of Jalisco and Colima. With $3259$ students coming from public high schools in the state of Jalisco, and $2814$ students from public high schools the state of Colima, we have a baseline sample of N $\sim{6073}$. We follow these students throughout the school year 2018-2019 and collect survey data in three different occasions.\Cref{tab:Descriptive-EduAspirations} shows educational aspirations, our analysis variable, in the three moments of the data collection. Note that we observe attrition after each collection period. We conduct an attrition analysis in \cref{subsection:attrition} and confirm that this does not pose a problem for our analysis.

\begin{table}[h!]
\centering
  \resizebox{1\textwidth}{!}{\begin{minipage}{\textwidth}
  \begin{center}
\caption{Educational Aspirations}
\label{tab:Descriptive-EduAspirations}

\hspace*{-0.5cm}\begin{tabular}{@{}cccccc@{}}

\toprule
\multicolumn{1}{l}{} & High school & Vocational Training & Undergraduate & Masters & \multicolumn{1}{l}{PhD}  \\ 
\midrule
\multicolumn{1}{l}{Baseline}  & \multicolumn{1}{l}{25} & \multicolumn{1}{l}{422} & \multicolumn{1}{l}{554}  & \multicolumn{1}{l}{1005} & 4066 \\ 
\multicolumn{1}{l}{Intervention} & \multicolumn{1}{l}{21} & \multicolumn{1}{l}{71} & \multicolumn{1}{l}{998} & \multicolumn{1}{l}{1061} & 2930 \\ 
\multicolumn{1}{l}{Follow up} & \multicolumn{1}{l}{23} & \multicolumn{1}{l}{58}  & \multicolumn{1}{l}{1009} & \multicolumn{1}{l}{1065} & 2541 \\ \bottomrule
\end{tabular}
\end{center}
\end{minipage}}
\end{table}

\paragraph{}
The baseline measure of our variable of interest, educational aspirations, exhibits a monotonic trend. Students' aspirations jump from  high school to vocational training and jump again at the Ph.D. level. The trend suggests students have very high aspirations, but there is a non-trivial number of students who do not aspire to a higher education degree. In general, this should not be a worrying trend unless the reason behind the choices originates from self-doubt or misinformation about the economic returns to a higher education degree \citep{jensen2010perceived, attanasio2014education}. Also noteworthy is the fact that the first jump (vocational training) disappears, and the second jump attenuates during the intervention and follow-up periods, in favour of undergraduate and master levels. In addition, with 67\% of baseline students aspiring to a doctoral degree, there may be a prevalent notion of increasing returns to education, when the reverse is true in Mexico \citep{caamal2017decreasing}. We denote student $i$'s educational aspirations at time $t$ with $y_{it}$, where $t = \{0,1,2\}$ corresponds to the data collection periods baseline [0], intervention [1] and endline or follow-up [2]. From $y_{it}$ we compute two outcome variables $y_{it}^c$ and $y_{it}^{dc}$. First, $y_{it}^c = \llbracket y_{it} \neq y_{it_{-1}} \rrbracket$ indicates whether student $i$'s educational aspirations are different at time $t$ with respect to time $t-1$; and $y_{it}^{dc}$ indicates whether aspirations decreased, increased or remained the same at time $t$ with respect to time $t-1$, such that:

$y_{it}^{dc} = \begin{cases} ~~1 & \text{if aspirations increased at time $t$ with respect to time $t-1$}\\ ~~0 & \text{if aspirations remained the same at time $t$ with respect to time $t-1$} \\ -1 & \text{if aspirations decreased at time $t$ with respect to $t-1$} \end{cases}$ 
\bigskip

\paragraph{}
\Cref{tab:Descriptive-IncAspirations} shows descriptive statistics of income aspirations. We consider income aspirations for our understanding of the general aspirational framework, but not as an outcome variable. Both the median and the mean student's income aspirations increase in the middle of the school year, coinciding with the intervention, but go back to pre-intervention numbers by the end of the school year.
\begin{table}[h!]
\centering
\begin{threeparttable}
\caption{Income Aspirations}
\label{tab:Descriptive-IncAspirations}

\hspace*{-0.5cm}\begin{tabular}{@{}ccccccc@{}}
\toprule
\multicolumn{1}{l}{} & Min. & 1st Quartile & Median & Mean & 3rd Quartile & \multicolumn{1}{l}{Max} \\ 
\midrule
\multicolumn{1}{l}{Baseline}  & \multicolumn{1}{l}{1000} & \multicolumn{1}{l}{10000} & \multicolumn{1}{l}{20000}  & \multicolumn{1}{l}{23007} & \multicolumn{1}{l}{30000} & \multicolumn{1}{l}{100000}\\ 
\multicolumn{1}{l}{Intervention} & \multicolumn{1}{l}{1000} & \multicolumn{1}{l}{15000} & \multicolumn{1}{l}{22000} & \multicolumn{1}{l}{28403}& \multicolumn{1}{l}{36000} & \multicolumn{1}{l}{100000} \\ 
\multicolumn{1}{l}{Follow up} & \multicolumn{1}{l}{1000} & \multicolumn{1}{l}{10000}  & \multicolumn{1}{l}{20000} & \multicolumn{1}{l}{23645} & \multicolumn{1}{l}{30000} & \multicolumn{1}{l}{100000} \\ \bottomrule
\end{tabular}
\begin{tablenotes}
      \scriptsize
      \item[1]Note: We restrict the minimum and maximum thresholds for income aspirations, from one fourth of the monthly minimum wage (USD 250) to the monthly wage of the top 1 per cent earners in the country.
      \item[2]Quantities in the table are expressed in Mexican Pesos (MXN).
    \end{tablenotes}
\end{threeparttable}
\end{table}

\paragraph{}
It is important to mention that, despite the different shift dynamics, we still observe a small but positive spearman correlation ($\rho$ = .137) between educational and income aspirations in the baseline sample, such that the higher the educational aspirations, the higher the income aspirations. This positive correlation slightly increases at mid and endline points ($\rho$ = 0.148), suggesting that income considerations are relevant for students when thinking about their education, and consistent with studies that find that subjective expectations about income are determinants of educational choices \citep[e.g.][]{attanasio2014education}.

\begin{figure}[h]
\begin{center}
\caption{Self-perceived social class of student: WVS \& GCN}
\label{figure:socialclass}
\includegraphics[width=\textwidth]{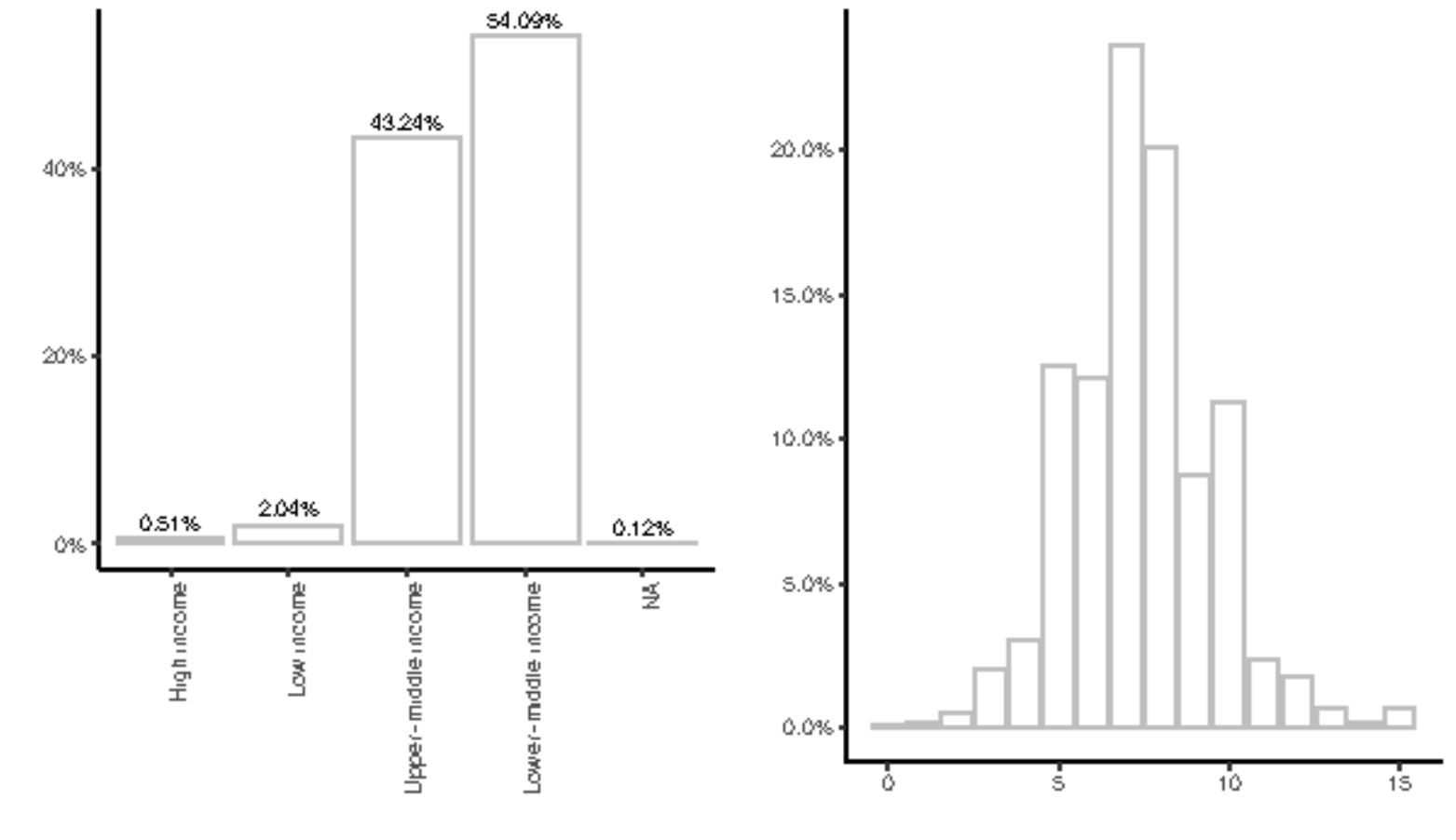}
\end{center}
\end{figure}

\begin{figure}[h]
\begin{center}
\caption{Barriers to achieving aspired educational degree}
\label{figure:edaspobstacle}
\includegraphics[width=0.8\textwidth]{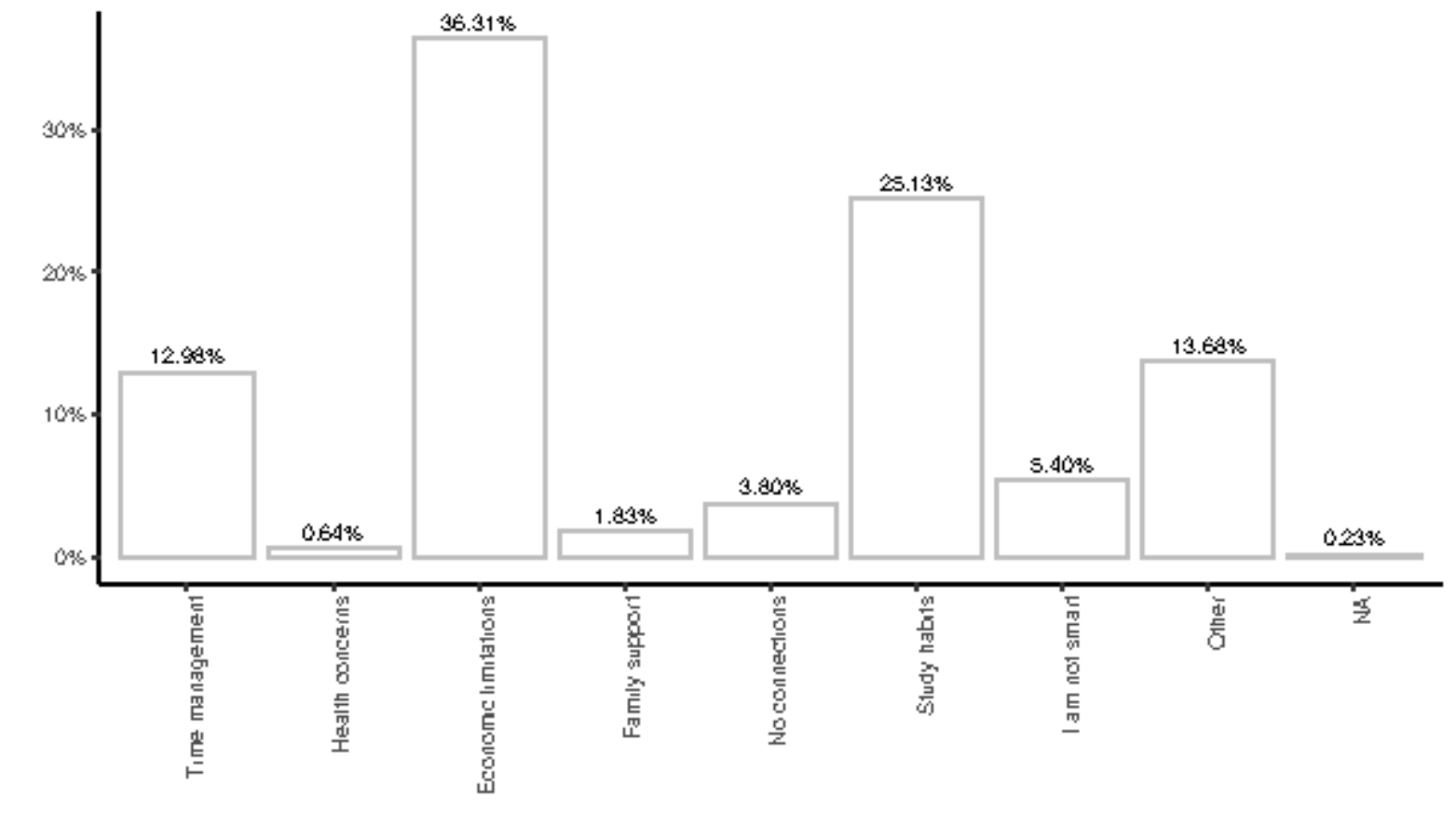}
\end{center}
\end{figure}

\paragraph{\textbf{Socioeconomic data:}}
We measure the socioeconomic status of students' through a subjective appraisal of their social class, as suggested by the World Values Survey [WVS]. The WVS question is a 4 level factor variable that allows students to identify themselves as low, lower-middle, upper-middle, or high income people. To allow for more variance in the measure, we [GCN] design a similar question that asks students to rate their socioeconomic status on a scale from 0 to 15, where zero means low, and fifteen means high income. \Cref{figure:socialclass} shows the distribution of both variables. The WVS factor variable indicates that the vast majority of surveyed students consider themselves to be lower and upper-middle income. The GCN interval variable exhibits a normal distribution of perceived income. With  97.33\% of the sample self-identifying themselves as average, any income based statistical inference is therefore to be taken with caution. Our sample seems to represent only middle-income students. Interestingly, 36.31\% of the sample also indicate that the biggest obstacle to achieving their aspired educational degree is economic limitations (see \cref{figure:edaspobstacle}). Therefore, there exists a subsample of students, identifying themselves as middle-income, for whom the economic elements emphasized in the video-intervention are relevant.

\paragraph{\textbf{Psychological data:}}

\begin{figure}[h]
\begin{center}
\caption{Histograms of psychological variables (average scores)}
\label{figure:psych}
\includegraphics[width=0.8\textwidth]{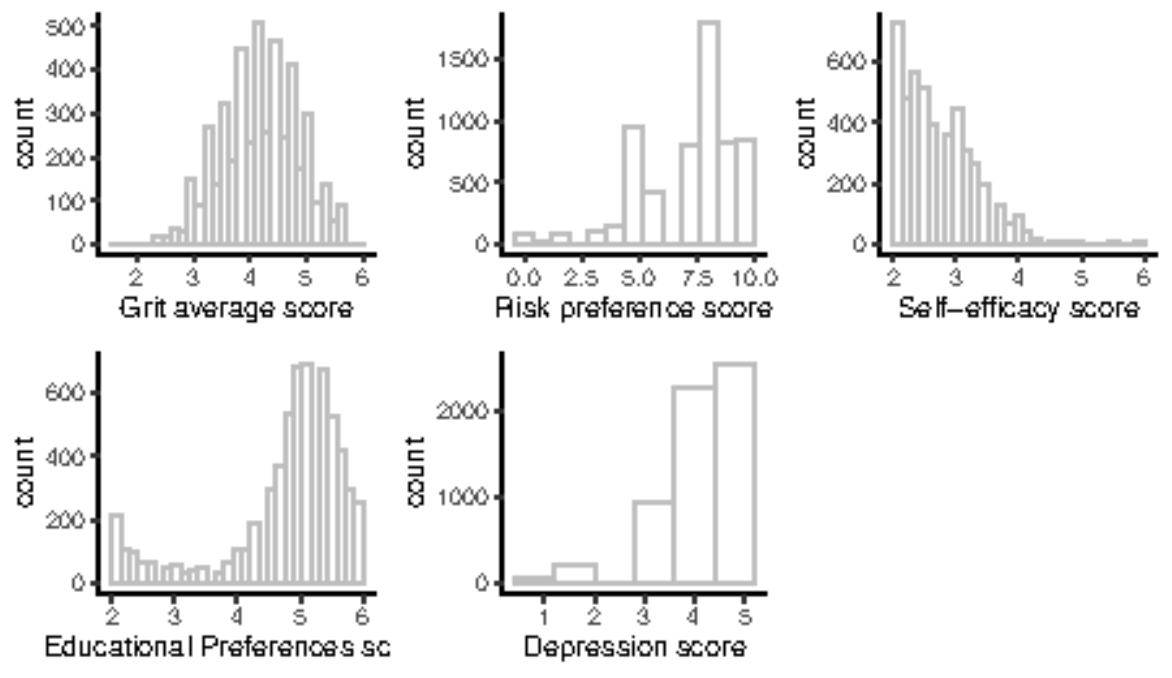}
\end{center}
\end{figure}

The distribution of psychological scores, in \cref{figure:psych}, shows some interesting dynamics. Grit concentrates the majority of observations in the middle values. The depression score exhibits a ladder-type behaviour with bunching in the higher scores. This is consistent with the notion that adolescents are susceptible to extreme emotions, and more likely to experience negative emotions \citep{verma1999adolescents}.This is also consistent with the observed self-efficacy score, with most values concentrating in the lower self-efficacy levels (little trust in one's own abilities). Finally, educational preferences [or how important a student believes education is for future success] are concentrated on the higher values of the 5 item score. This is in line with the trend we observe in educational aspirations, and may be a reflection of a strong belief that participation in higher education is a way out of the working class \citep{gregorutti2011commercialization}.

\paragraph{\textbf{Network data:}}
\Cref{tab:summary_network_vars} presents summary statistics for the peer network variables. All parameters are estimated as features of the nodes (students), except for `global transitivity'\footnote{We use this feature in the matching strategy, but refer to it as the `global clustering coefficient'.}, which is a feature of the school network. Similarly, `betweenness centrality' is estimated using the school as the reference graph, as opposed to the overall sample graph. 

\begin{table}[!htbp] \centering 
  \caption{Summary Statistics for Network Variables} 
  \label{tab:summary_network_vars} 
\scalebox{1}{\begin{tabular}{@{\extracolsep{5pt}}lccccccc} 
\\[-1.8ex]\toprule 
 
Statistic & \multicolumn{1}{c}{N} & \multicolumn{1}{c}{Mean} & \multicolumn{1}{c}{St. Dev.} & \multicolumn{1}{c}{Min} & \multicolumn{1}{c}{Pctl(25)} & \multicolumn{1}{c}{Pctl(75)} & \multicolumn{1}{c}{Max} \\ 
\midrule \\[-1.8ex] 
transitivity(local) & 6,051 & 0.203 & 0.111 & 0.000 & 0.138 & 0.250 & 1.000 \\ 
betweenness & 6,073 & 0.025 & 0.039 & 0.000 & 0.002 & 0.031 & 0.362 \\ 
strength & 6,073 & 35.784 & 17.928 & 1 & 23 & 45 & 165 \\ 
degree & 6,073 & 10.034 & 4.582 & 1 & 7 & 13 & 39 \\ 
Indegree & 6,073 & 4.919 & 2.520 & 0 & 3 & 6 & 17 \\ 
Outdegree & 6,073 & 5.115 & 2.860 & 1 & 3 & 6 & 23 \\ 
transitivity(global) & 6,073 & 0.436 & 0.059 & 0.330 & 0.385 & 0.471 & 0.579 \\ 
\bottomrule \\[-1.8ex] 
\end{tabular}} 
\end{table} 

\paragraph{\textbf{Experimental subsample:}}
\label{subsubsection:ExperimentalSubsample}
In \cref{subsection:ExperimentalDesign} we define $A^s$ as the set of students in each school that satisfy the treatment eligibility criteria. Therefore, the subsample $n^A = 1556$ from the total student population $N = 6073$, contains only students that are eligible for treatment across all schools; i.e. comparable students between treated and non-treated schools. \Cref{tab:ycd-by-treatment} is a contingency table of treated and control students from the subsample $n^A$ across values of $y^{cd}_i$, from baseline to intervention, and baseline to endline periods. That is, the short and longrun changes of educational aspirations (within the school year). Note that we perform the causal evaluation of the video-intervention on this subsample, but study \textit{causal} spillover effects of the video-intervention on the full sample. The contingency table shows that there is change in aspirations in both groups of students, but treated students seem to change their aspirations in larger numbers than in control schools. With $p = 0.000008$, and $p = 0.09817$ respectively, a Fisher's Exact Test indicates that the movement in aspirations from these two groups are different from each other, i.e. drawn from independent distributions. Our claim is that the difference in the movement is due to our video-intervention. To test it, we estimate the causal effects with non-linear regressions in \cref{section:results_discussion}.

\begin{table}[!htbp] \centering 
  \caption{Change of educational aspirations between treated and control students, from baseline to intervention*, and baseline to endline.**} 
  \label{tab:ycd-by-treatment} 
\scalebox{1}{\begin{tabular}{@{\extracolsep{5pt}}lcccc} 
\\[-1.8ex]\toprule 
 
&\multicolumn{1}{c}{Decrease aspirations (-1)} & \multicolumn{1}{c}{Aspirations do not change (0)} & \multicolumn{1}{c}{Increase aspirations (1)}\\ 
\midrule \\[-1.8ex] 
Treated* & 229 & 419 & 114\\
Control* & 164 & 528 & 102\\
\midrule \\[-1.8ex] 
\\
\midrule \\[-1.8ex]
Treated** & 183 & 482 & 97\\
Control** & 167 & 543 & 84\\
\bottomrule \\[-1.8ex]
\end{tabular}} 
\end{table}

\paragraph{\textbf{Power and sample size determination:}}
To determine the number of schools and participating students in the study we follow three strategies. First, for Colima, we observed the size of grade 12 cohorts of all public high schools in the state, and dropped those schools whose number was in the lowest quartile of the sample. Second, for Jalisco, we randomly selected a similar number of public schools and made sure that grade twelve cohorts were neither smaller nor too large from those in Colima, with at least 120 students per cohort and no more than 500. Our final school sample consists of 45 public high schools of which 22 are in Colima and 23 in Jalisco. The total grade 12 population, our target group, consists of 6,073 students that willingly participate in the study. To determine the number of pupils that receive the intervention within each treatment school, we use a deterministic algorithm similar to \citepos{paluck2016changing}.\footnote{\citet{paluck2016changing} select a fixed 15 per cent of each school's population, we vary up to 25 per cent. The school proportions can be found in \cref{tab:school-prop}.} Our final sample consists of 762 treated students from treated schools, and 794 treatment-eligible students from non-treated schools that serve as our control students, such that $n^A = 1556$. 

\paragraph{}
We compute the experiment's power score with a post-hoc power calculation using R's \textit{pwr} package for the difference in proportions of our binary outcome variable, and the \textit{DescTools} package to determine the power of our multinomial outcome. For the binary outcome $y_{it}^c$, we obtained a cohen's H value of 0.236, also known as the detectable effect size \citep{cohen2013statistical}, given our treatment group size of $n_1$ 762 and control group size of $n_2$ 794, at a significance level $\alpha$ = 0.001, and power of 0.91. Meaning, we can confidently report small, medium and large effect sizes for $y_{it}^c$ given our current sample size. For the multinomial outcome $y_{it}^{dc}$, we estimated a minimum desired sample size of 383.668 subjects per treatment arm given a significance level of  $\alpha$ = 0.05, and a minimum desired sample size of 669.95 per treatment arm given a significance level of $\alpha$ = 0.01. \Cref{tab:multinomial_ci} provides the estimtes of each category in the outcome variable and, using the Sison and Glaz method, two-sided simultaneous confidence intervals. The desired minimum sample size $n^*$ was computed with the formula $t^2 / 4(E)^2$; where $E$ is the difference in CIs. In the case of outcome $y_{it}^{dc}$, our study is also sufficiently powered. 

\begin{table}[!htbp] \centering 
  \caption{Simultaneous confidence intervals for multinomial proportions} 
  \label{tab:multinomial_ci} 
\scalebox{1}{\begin{tabular}{@{\extracolsep{4pt}} cccc} 
\toprule \\[-1.8ex] 
 & $P$ & Lwr.ci & Upr.ci \\ 
\midrule \\[-1.8ex] 
~0 & $0.609$ & $0.578$ & $0.639$ \\ 
-1 & $0.253$ & $0.222$ & $0.283$ \\ 
~~1 & $0.139$ & $0.109$ & $0.169$ \\ 
\bottomrule \\[-1.8ex] 
\end{tabular}}
\end{table}

\subsection{Estimation approach}
\label{subsecton:Estimation approach}
\subsubsection{Causal Effects under Peer Influence}
\label{subsubsection: Causality_and_peers}

Traditionally, causal inference analysis has focused on individual response to treatment. That is, we assume that the outcomes of a person are unaffected by the treatment status of another. But in the presence of social interactions, this assumption doesn't hold. It fails to account for two types of effects: within-group effects, i.e. a change in the magnitude of treatment effects induced by social interactions between treated individuals \citep{manski2013identification}; and between-group effects (spillovers), i.e. a change in the magnitude of the effect induced by social interactions between treated and non-treated individuals \citep{dieye2014accounting,hudgens2008toward, aronow2017estimating}.
    
\paragraph{}
In response to this concern, we propose a hierarchical experimental design where student selection is semi-aleatory and stratified by gender and betweenness centrality. In addition, we use  above-median strength as a selection cut-off for intervention-eligible students.
From here on, we use $i$ to refer to the student index and $s$ to refer to the school index. We define $\bm{z}$ such that $z_i \in \{0,1\}$ denotes whether student $i$ has satisfied the individual selection criteria or not, and $z^s \in \{0,1\}$ denotes whether school $s$ was selected for treatment or not. $\bm{G}_i$ is the row for student $i$ in the network adjacency matrix, such that $G_{ij}$ takes the value 1 if students $i$ and $j$ are friends, and 0 otherwise. 

\paragraph{}
We adopt the peer effect identification strategy in \citet{dieye2014accounting}, and assume that student $i$'s outcome $y$ is a function of her friends' mean outcomes $\bm{G}_{i}y$, her treatment status $z_i$, and her friends' treatment status $\bm{G}_{i}\bm{z}$:

\begin{equation} \label{eq:1Treatment_status_and_peer_effects} 
    y_i^* = \alpha + \beta(\bm{G}_iy) + \gamma z_i + \delta(\bm{G}_i\bm{z}) + u_i
\end{equation}

Where $y_i^*:= \{y^c_i, y^{dc}_i\}$, $\bm{G}_iy$ refers to student $i$'s friends' mean educational aspirations and $\bm{G}_i\bm{z}$ is the proportion of her friends that are treated. $\alpha$ denotes a constant and $u_i$ is an error term.

\paragraph{}
To measure school spillovers, i.e. treatment diffusion in treated schools, we define the vector $z_j^s = z^s - z_i$, which indicates whether a non-eligible for treatment student $j$ belongs to a school selected for treatment or not.  We then follow the same model in \cref{eq:1Treatment_status_and_peer_effects}:

\begin{equation} \label{eq:2School_Spillovers}
   y_i^* = \alpha + \beta(\bm{G}_iy) + \gamma z_j^s + \delta(\bm{G}_i\bm{z}) + u_i
\end{equation}

In \cref{eq:2School_Spillovers} the treatment vector $z_j^s$ takes on the value of 1 if non-treated student $j$ belongs to a school that was selected for treatment and zero otherwise. We define the indicator as such to isolate the effect of the peer network spillover from the treatment effect that is potentially present in the set of students that satisfy the selection criteria $A^s$ in each school.

\paragraph{Covariate balance and inclusion:}
\Crefrange{eq:1Treatment_status_and_peer_effects}{eq:2School_Spillovers} are estimated without covariates, as is common in randomised experiments. We present a covariate list and balance analysis in the appendix, \cref{tab:cov-balance}. Despite 2 of the 16 covariates being unbalanced, we abstain from including them in the main non-linear models; \citet{bruhn2009pursuit} outline three major reasons why covariate balance analyses are not required (irrespective of what they might show), including that they are conceptually redundant as you already know that the imbalance occurred by chance, i.e. there is no sorting \citep{altman1985comparability}, that the focus on the imbalance of a variable that is uncorrelated with the outcome of interest detracts from the main analysis of the model, and may decrease precision of the estimates \citep{green2011analyzing}, and that there is a large probability that these tests are being used improperly \citep{schulz2002allocation}.\footnote{We nevertheless include all the non-linear models with the balanced and unbalanced covariates in \cref{subsection:cov-balance} as robustness checks. The results corroborate the conclusions drawn from the main models, albeit a small difference in magnitudes of the estimated coefficients of interest.} \citet{green2011analyzing} argue that covariate adjustment will lead to increased precision of estimates and predictive power of the average treatment effect if and only if, the sample is large enough\footnote{\citet{freedman2008regression} argues that the minimum sample size is $n_{min}=1000$, whereas \citet{green2011analyzing} say $n_{min}=20$.}, covariates are not influenced by the treatment status, and covariates have a strong predictive power of the outcome variable. If these three conditions do not apply, covariate adjustment will instead induce bias to the estimates and decrease precision. 


\subsubsection{A comment on the identification of peer effects}

Treatment randomization ensures that any potential effects originating from unobserved confounds, environmental shocks, or sorting are balanced between experimental groups $\bm{z}$. Thus, the estimated parameters $\gamma z_i$, $\gamma z_j^s$ and $\delta(\bm{G_iz})$ in \cref{eq:1Treatment_status_and_peer_effects} \& \cref{eq:2School_Spillovers} can be considered causal effects in our experimental design. However, parameter $\beta(\bm{G}_iy)$ is subject to what \citet{manski1993identification} calls the reflection problem. That is, student $i$'s outcome $y_i^*$ may be affected by student $j$'s outcome $y_j^*$, but $j$'s outcome $y_j^*$ may also be affected by $i$'s outcome $y_i^*$. \citet{bramoulle2009identification} propose a solution that \citet{dieye2014accounting} describe as an instrumental variable that stems from the heterogeneity in the social network architecture: intransitivity. As long as there exist some open friendship triangles such that $i$ is friends with $j$, $j$ is friends with $k$ and $k$ is \emph{not} friends with $i$, $i$'s treatment status should only affect $k$'s behaviour through its effect on $j$. In practice, this amounts to testing whether the student friendship graph $G$, its identity matrix $I$ and its square $G^2$ are linearly independent\footnote{We determine this using the R packages Caret and Matrix, which provide the functions DetectLinearCombos and rankMatrix.}. We provide a table of open triangles per school graph $G^s$ and discuss linear independence results in \cref{subsection:peer-identification} and \cref{tab:school-intransitivity}. 

\paragraph{}
Furthermore, we previously propose vector $z^s_j$ as a way to capture school-wide treatment diffusion.  However, this strategy does not capture the compound effect, i.e. the effect of social interactions on the net effect of the intervention. A way of estimating this, as proposed by \citet{dieye2014accounting} is to compute the composite social effect using the estimated parameters from \cref{eq:1Treatment_status_and_peer_effects}:

\begin{equation} \label{Eq:4 Composite social effect} 
   \Delta^{student\_pop} = \frac{\gamma + \delta}{1-\beta}
\end{equation}

The composite social effect captures the net effect of the intervention by accounting for the treatment effect, and the endogenous and exogenous peer effects. 

\section{Results and Discussion}
\label{section:results_discussion}
\subsection{Change in educational aspirations}
\label{subsection:results:chance in educ asp}
\paragraph{General model:}
\Cref{eq:1Treatment_status_and_peer_effects} in \cref{subsubsection: Causality_and_peers} is estimated using a logistic regression model. We present results as Average Marginal Effects for parameter interpretation purposes.  \Cref{tab:AME:logistic_yc} shows regression results for $y_i^c$, change in educational aspirations, where treatment status $z_i \in\ \{0,1\}$ indicates whether the student was invited to the video screening or not. In the first moment [01], from baseline to intervention, being invited to the screening increases the probability of changing your aspirations by 12 percentage points (p $<$ 0.001). Peer effects, in the form of the average educational aspirations of student $i$'s friends, are also statistically significant (p $<$ 0.001) and they increase the probability to change aspirations by 20 percentage points. On the other hand, peer effects in the form of student $i$'s friends' treatment status are not statistically significant. These results are in line with what is expected given the intervention implementation process. Students were surveyed the same day as selected students watched the video-intervention, and would therefore have not had much time to interact with their peers/friends after the fact. Interestingly, between the second and third moments, the treatment status stops being statistically significant and the magnitude of its coefficient decreases. In the second moment [21], between the intervention and the endline period, the only statistically significant parameter is that of the proportion of student $i$'s treated friends at p $<$ .001, and a coefficient of 0.12 pp. In the third moment [20], between baseline and endline, the change in aspirations is completely driven by social interactions. That is, the average aspirations of student $i$'s friends increase the probability of changing her aspirations by 18.5 percentage points at p $<$ .001, and the proportion of student $i$'s treated friends increases the probability to change her aspirations by 15 percentage points, also at p $<$ .001. These results coincide with those of other successful behavioural interventions; first, there is a peak in the effect of the treatment status directly after the intervention \citep{allcott2014short}, which attenuates over time. Second, both directly post intervention and in the long run (where the long run is measured as baseline to endline, 02), the effects of the treatment are primarily driven by social interactions, with the fraction of treated friends consistently influencing behaviour \citep[similar to][]{paluck2016changing,dieye2014accounting}. To further investigate the mechanism through which social interactions influence behaviour, we include in the model a variable that captures dynamic peer effects in \cref{tab:mechlogistic_yc} of \cref{subsection:mechanism-reg}. The variable expresses whether student i's friends revise their aspirations at time $t$ with respect to time $t-1$. In the model, dynamic peer effects highly correlate with the fraction of treated friends, as they are more likely to revise their aspirations. Unsurprisingly, it also absorbs the effect of own treatment status. This result is consistent with our findings in the main model, and reinforces the narrative that social multipliers or social stabilisers can only exist if the treatment status induces behavioural change.

\paragraph{Composite social effect:}
The composite social effect for \cref{eq:1Treatment_status_and_peer_effects}, or the net causal effect of the intervention under social interactions, is 0.18 pp for the first moment [10]. For the second [21] and third [20] moments, the respective coefficients are 0.17 pp  and 0.23 pp. The composite social effect is homologous to the ATT of a randomised experiment when observational units are isolated, and therefore interactions across experimental groups and within experimental groups do not affect the treatment outcome. I.e. the Stable Unit Treatment Value Assumption (SUTVA) holds. From \cref{tab:AME:logistic_yc} we can directly observe that, when it comes to social or behavioural interventions, subject interactions within experimental groups do affect the treatment outcome. Therefore, any experimental analysis of a setup where sociability cannot be contained would result in conservative estimates of the net effect of the treatment on the outcome (and a violation of SUTVA) \citep{sinclair2012detecting,aronow2017estimating}. Comparing coefficient results between the composite social effect and the treatment effect of the first model we corroborate the statement above; the true net effect of the intervention is larger when accounting for endogenous and contextual peer effects.

\begin{table}[!htbp] \centering 
  \caption{Average Marginal Effects: Logistic regression results for three moments [$z_i$]} 
  \label{tab:AME:logistic_yc} 
\begin{tabular}{@{\extracolsep{5pt}}lccc} 
\\[-1.8ex]\toprule 

 & \multicolumn{3}{c}{\textit{Dependent variable:}} \\ 
\cmidrule(l){2-4} 
\\[-1.8ex] & Educ. aspirations $y^c$ 10 & Educ. aspirations $y^c$ 21 & Educ. aspirations $y^c$ 20 \\ 
\\[-1.8ex] & (1) & (2) & (3)\\ 
\midrule \\[-1.8ex] 

Treated    & $0.119^{***}$ & $0.045$       & $0.042$       \\
               & $(0.028)$     & $(0.024)$     & $(0.027)$     \\
PeerEffects(avg)          & $0.201^{***}$ & $0.034$       & $0.185^{***}$ \\
               & $(0.031)$     & $(0.027)$     & $(0.030)$     \\
PropTreatFr         & $0.026$       & $0.120^{***}$ & $0.148^{***}$ \\
               & $(0.020)$     & $(0.019)$     & $(0.021)$     \\
\midrule
Num. obs.      & $1251$        & $1251$        & $1251$        \\
Log Likelihood & $-803.862$    & $-674.947$    & $-759.854$    \\
Deviance       & $1607.725$    & $1349.894$    & $1519.709$    \\
AIC            & $1615.725$    & $1357.894$    & $1527.709$    \\
BIC            & $1636.251$    & $1378.421$    & $1548.236$    \\
\bottomrule
\multicolumn{4}{l}{\scriptsize{Note: $^{***}p<0.001$; $^{**}p<0.01$; $^{*}p<0.05$}}\\
\multicolumn{4}{l}{\scriptsize{Reported standard errors are HC3}}
\end{tabular} 
\end{table}

\paragraph{School spillovers:}
While the composite social effect allows us to better understand the role of social interactions on the treatment effect, it does not directly measure spillovers to the student population. We measure school spillovers by estimating \cref{eq:2School_Spillovers}, where $$z_j^s$$ is a school treatment vector such that $A^s \centernot\in z_s$. In \cref{tab:eq1spillovers} we compare non-treated students from treated schools vs. non-eligible students from non-treated schools. By removing treated students from the school-treatment assignment vector, we guarantee that the observed treatment we can disentangle the effect of the treatment on the outcome from the effect of the treatment spillover. This is directly reflected on the coefficient of the treatment assignment vector, which is both small in magnitude and not statistically significant for any of the moments. However, we note that both types of peer effects are statistically significant throughout the school year, with the proportion of treated friends consistently significant at p $<$ 0.001. Much like what we observed from the general model, in the long run [02] it is the proportion of treated friends that carry the effect, increasing the probability of changing aspirations by 7.9 percentage points for non-treated students in treated schools. We can conclude that the second-order peer effects are non-negligible and, in fact, in a similar order of magnitude as the first-order peer effects.

\begin{table}[!htbp] \centering 
  \caption{Average Marginal Effects: Logistic regression results for three moments [$z_j^s$]} 
  \label{tab:eq1spillovers} 
\begin{tabular}{@{\extracolsep{5pt}}lccc} 
\hline \\[-1.8ex] 
 & \multicolumn{3}{c}{\textit{Dependent variable:}} \\ 
\cline{2-4} 
\\[-1.8ex] & Educ. aspirations $y^c$ 10 & Educ. aspirations $y^c$ 21 & Educ. aspirations $y^c$ 20 \\ 
\\[-1.8ex] & (1) & (2) & (3)\\ 
\hline \\[-1.8ex] 
 
 TreatedSchool & $0.015$       & $0.030$       & $0.010$       \\
                  & $(0.015)$     & $(0.016)$     & $(0.013)$     \\
PeerEffects(avg)             & $0.240^{***}$ & $0.195^{***}$ & $0.038^{*}$   \\
                  & $(0.017)$     & $(0.018)$     & $(0.015)$     \\
PropTreatFr            & $0.027^{**}$  & $0.123^{***}$ & $0.079^{***}$ \\
                  & $(0.009)$     & $(0.010)$     & $(0.008)$     \\
\hline
Num. obs.         & $3777$        & $3777$        & $3777$        \\
Log Likelihood    & $-2252.190$   & $-2250.817$   & $-1944.009$   \\
Deviance          & $4504.381$    & $4501.634$    & $3888.019$    \\
AIC               & $4512.381$    & $4509.634$    & $3896.019$    \\
BIC               & $4537.328$    & $4534.580$    & $3920.965$    \\
\hline
\multicolumn{4}{l}{\scriptsize{Note: $^{***}p<0.001$; $^{**}p<0.01$; $^{*}p<0.05$}} \\
\multicolumn{4}{l}{\scriptsize{Reported standard errors are HC3}}
\end{tabular} 
\end{table} 

\subsection{Direction of change in educational aspirations}

We estimate \cref{eq:1Treatment_status_and_peer_effects} in \cref{subsubsection: Causality_and_peers} using a multinomial logistic model for $y^{cd}_i$ with $z_i$ and present results using Relative Risk Ratios. As before, we present three models corresponding to the three moments of the experiment but we do not compute the composite social effect from \cref{Eq:4 Composite social effect} nor school spillovers from \cref{eq:2School_Spillovers}. Our interest with this formulation is to focus on the direction of the effect, i.e. whether aspirations decrease or increase, and whether the parameters mimic the dynamic from our previous analysis, from which we can infer positive second-order network effects.

\begin{sidewaystable}[!htbp] \centering 
  \caption{Relative Risk Ratios: Multinomial model for 3 moments [$z_i$]} 
  \label{tab:multinomial1} 
\begin{tabular}{@{\extracolsep{5pt}}lcccccc} 

\hline \\[-1.8ex] 
 & \multicolumn{6}{c}{\textit{Dependent variable:}} \\ 
\cline{2-7} 
\\ & -1 & 1 & -1 & 1 & -1 & 1 \\ 
\\ & $y_i^{dc}$(base 10) & $y_i^{dc}$(base 10) & $y_i^{dc}$(long 20) & $y_i^{dc}$(long 20) & $y_i^{dc}$(end 21) & $y_i^{dc}$(end 21)\\ 
\hline \\[-1.8ex] 
Treated & 1.777$^{***}$ & 1.491$^{***}$ & 1.213$^{***}$ & 1.251$^{***}$ & 1.070$^{***}$ & 1.534$^{***}$ \\ 
& (0.243) & (0.318) & (1.71e-1) & (2.80e-1) & (0.189) & (0.272)\\ & & & & & & \\ 
PeerEffects(avg) & 0.417$^{**}$ & 55.649$^{***}$ & 0.412$^{**}$ & 69.518$^{***}$ & 1.070$^{***}$ & 1.349$^{***}$ \\ 
& (0.0830) & (16.3) & (8.51e-2) & (2.30e+1) & (0.212) & (0.257)\\ & & & & & & \\ 
PropTreatFr & 1.209$^{***}$ & 0.911$^{***}$ & 1.909$^{***}$ & 2.178$^{***}$ & 1.720$^{***}$ & 2.223$^{***}$ \\ 
& (0.121) & (0.139) & (2.14e-1) & (3.92e-1) & (0.241) & (0.332)\\ & & & & & & \\ 
\hline \\[-1.8ex] 

Akaike Inf. Crit. & 1,881.418 & 1,881.418 & 1,748.657 & 1,748.657 & 1,788.835 & 1,788.835 \\ 
\hline 
\multicolumn{4}{l}{\scriptsize{Note: $^{*}$p$<$0.1; $^{**}$p$<$0.05; $^{***}$p$<$0.01}} \\ 
\multicolumn{4}{l}{\scriptsize{Standard errors are computed using the Taylor series-based delta method}}\\ 
\end{tabular} 
\end{sidewaystable}

\paragraph{}
\Cref{tab:multinomial1} provides regression results for $y_i^{cd} \in \{-1,0,1\}$, direction of the change in educational aspirations. That is, do aspirations decrease or increase? Our reference category in the multinomial model is zero, or educational aspirations remain the same at time $t$ from time $t-1$. The treatment status $z_i \in \{0,1\}$ indicates whether the student was invited to the video screening o not. 

\paragraph{}
In the first moment, student $i$'s treatment status' odds of decreasing their aspirations as compared to not changing them, are 77\% higher (p $<$ 0.01),  and the odds of increasing their aspirations are 49\% higher (p $<$ 0.01). For each educational level increase in the average educational aspirations of student $i$'s friends, the odds of decreasing her aspirations are 58\% lower (p $< 0.05$), and the odds are 44.4\% lower for increasing her aspirations (p $<$ 0.01). With a higher proportion of treated friends we also observe higher odds (20\%) of decreasing aspirations, and lower odds (8\%) of increasing her aspirations; both coefficients are statistically significant at p $<$ 0.01. Put otherwise, a percentage point increase in the proportion of treated friends increases the odds of decreasing aspirations by 20\% for students with very high aspirations, ceteris paribus; and a percentage point increase in the proportion of treated friends decreases the odds of increasing her aspirations by 8\%, all else constant. As discussed in \cref{subsection:results:chance in educ asp}, social interactions between treated students are limited in the first moment, therefore this result may hint at expectation-based behaviour. That is, for those students who do change their aspirations, they may revise their aspirations taking into account that some fraction of their friends are (also) revising. Unfortunately, we cannot analyze this idea in detail with our data.  

\paragraph{}
For the second and third moments, all six estimated parameters remain strongly significant, and while effect sizes vary, they remain close to the estimated magnitudes for the first moment. There are two trends to highlight here; first, it seems like the peer effects that refer to the average aspirations of their friend group decrease the odds of both increasing and decreasing aspirations for the first moment [10] and in third moment [20], effectively acting as an anchor for change. This is not true for the second moment [21], or from intervention to endline, where the odds of increasing and decreasing aspirations are both higher. In the long run, however, the average aspirations of student $i$'s friends\footnote{Measured at $t_0$}, serve as a reminder not to deviate from a \say{predefined} friend consensus, as proposed in social learning models \citep{chandrasekhar2020testing}. Second, it is clear that the revision upward and downward coincides with the bimodal distribution observed in educational aspirations at baseline (see \cref{tab:Descriptive-EduAspirations}. In the video-intervention, students find out that educational degrees in Mexico exhibit decreasing returns; an undergraduate and a master degree will return higher salaries, but a Ph.D. will return a lower salary than a master degree. Students with low aspirations, are encouraged to revise their aspirations by the motivational and aspirational component of the video, as much as by the economic component of the video. Students with high aspirations (the majority of the sample), are encouraged to revise downward by the economic component of the video if, and only if their educational aspiration was based on the incorrect belief that the highest level of education possible would also provide the highest salary. In essence, what we observe from the multinomial regression results in \cref{tab:multinomial1} is the consequence of two types of students in the sample: the ca. 30\% of students that are burdened by economic limitations that may influence their educational aspirations, and students who have high educational aspirations with imperfect knowledge on the returns to education in Mexico. 

\section{Concluding remarks}

\paragraph{Limitations and future work:}
An important limitation of our study is that, at the time of writing, we are unable to collect data on student enrollment to analyse whether educational aspirations, particularly for the low-aspiration cum middle-aspiration subsample, translate into actual university enrollment behaviour. Moreover, we are also limited in our understanding of baseline information structures of Mexican adolescents. We did not collect data on what students believe graduate degrees, in particular a Ph.D. degree, will return salary-wise, nor did we collect information on students' awareness of salary differentials across educational disciplines. While we are able to infer the first element with the data we do have, the second missed element means that we cannot explore heterogeneity in educational specialisation preferences. 
Future studies on this topic would benefit from studying different types of peer-networks other than friendship, and the development of adolescent-specific measurements of subjective socioeconomic items, such as social class, gender diversity and ethnicity. Finally, to make stronger claims on the external validity of these type of interventions, a larger experimental sample is needed; a larger sample both in terms of block size (say, number of students and schools within a state), and number of blocks (increase the number of states involved).

\paragraph{Conclusion:}
Choosing to invest in their education is one of the most important decisions an adolescent has to make. Not only is higher education associated with social \citep{lochner2011non, heckman2018nonmarket} and economic \citep{psacharopoulos2018returns} benefits, but it also leads to improved social interactions and social trust \citep{oreopoulos2011priceless}. Choosing to invest in higher education is therefore good for young people and for the communities they are a part of. In this context, we study the educational decision-making process of adolescents in Mexican high schools, focusing on their educational aspirations, an established predictor of realised years of schooling (see \cref{subsection:malleabilityofasp}).  

\paragraph{}
We implemented a multilevel networked experiment in a sample of public high schools from the Northwestern region of Mexico. In the experiment, a semi-random subsample of students from selected high schools was invited to participate in the screening of a video-intervention. The 15 minute video exposed students to role models and information about the economic returns to education in the country. In line with the literature, we find that educational aspirations are malleable, that adolescents have partial and/or incorrect knowledge about the economic returns to education, and that peer social networks can be the catalyst for behavioural change. Our main contributions, however, are at the intersection of these three elements. First, to the best of our knowledge, we are the first study to establish a causal link between peer networks and aspirations change, although this connection has been discussed before \citep[e.g.][]{mani2019social, bernard2014future, gardineraspirations}. We also provide an empirical example of how to utilise peer networks to increase the reach of educational policies or programmes, in particular low cost technology-driven interventions. Second, we contribute to the evaluation literature by providing a clear instance in which the outcomes of an individual are affected by the treatment status of another, despite there being no interference between experimental groups. Our study highlights the importance of network data to circumvent estimation errors that may arise from a SUTVA violation in the context of social or schooling programmes. Moreover, we show that the mechanism through which a belief-adaptation intervention lasts is sociability. That is, the cognitive process of learning from an exogenous shock is stabilised when information is socialised. As a final note, our two state sample selection allows us to comment on the external validity of our experiment. As mentioned in \cref{section:intro}, these two states are dissimilar in various ways. Nonetheless, treatment and peer effect results are similar\footnote{We ascertain this claim by looking at the coefficient and statistical significance of the state binary covariate in \cref{subsection:cov-regress}. We further estimated the models in state subsamples and confirmed that the treatment effect remained significant. However, due to the small sample sizes resulting from the state subsampling strategy, we refer to the covariate-controlled regressions for statistical inference.}. This means that, overall, a network driven, low cost technology approach to aspirations in education is appropriate in two different Mexican contexts. Note, however, that the state binary (with Jalisco as baseline) in the first model, change in aspirations, is statistically significant in the long run [20] for the experimental subsample and in the short run [10] for the school subsample, albeit with very small effect sizes. This is an indication that the treatment effect (an indeed, experimental strategy) may be stronger in said state. We can therefore make three important claims: i. a role-model/aspirational video-intervention is effective in different (Mexican) contexts; ii. peer effects are consistently present across environments; iii. a social network approach is generally applicable in different contexts, albeit with differing strength.

\pagebreak
\bibliographystyle{abbrvnat}
\bibliography{library}

\pagebreak
\section{Appendix}
\subsection{School attributes}
\label{subsection:School-attributes}

\setlength{\arrayrulewidth}{0.1mm}
\setlength{\tabcolsep}{12pt}
\renewcommand{\arraystretch}{2.0}
\begin{ThreePartTable}
    \begin{TableNotes}
    \scriptsize
        \item[1]Note: Bachilleratos belong to the state of Colima and Preparatorias belong to the state of Jalisco.
    \end{TableNotes}
\begin{longtable}{|p{3.8cm}|p{3.8cm}|} 
\caption{School-Matched Treatment Assignment} \label{tab:school-match} \\
\hline

\multicolumn{1}{l}{Treated Schools} & \multicolumn{1}{l}{Control Schools}  \\                 
\hline
\multicolumn{1}{l}{Preparatoria 20 vespertino}     & \multicolumn{1}{l}{Preparatoria 13 vespertino}     \\
\multicolumn{1}{l}{Bachillerato 20}                & \multicolumn{1}{l}{Bachillerato 4}                 \\
\multicolumn{1}{l}{Preparatoria 5 vespertino}      & \multicolumn{1}{l}{Preparatoria 5 matutino}        \\
\multicolumn{1}{l}{Bachillerato 2}                 & \multicolumn{1}{l}{Bachillerato 16}                \\
\multicolumn{1}{l}{Preparatoria 17 matutino}       & \multicolumn{1}{l}{Preparatoria 17 vespertino}     \\
\multicolumn{1}{l}{Bachillerato 23}                & \multicolumn{1}{l}{Bachillerato 22}                \\
\multicolumn{1}{l}{Preparatoria 18 matutino}       & \multicolumn{1}{c}{Preparatoria Jalisco vespertino}\\
\multicolumn{1}{l}{Bachillerato 17}                & \multicolumn{1}{l}{Bachillerato 27}                \\
\multicolumn{1}{l}{Preparatoria 15 vespertino}     & \multicolumn{1}{l}{Preparatoria 15 matutino}       \\
\multicolumn{1}{l}{Bachillerato 10}                & \multicolumn{1}{l}{Bachillerato 6}                 \\
\multicolumn{1}{l}{Preparatoria 9 matutino}        & \multicolumn{1}{l}{Preparatoria 9 vespertino}      \\
\multicolumn{1}{l}{Bachillerato 30}                & \multicolumn{1}{l}{Bachillerato 3}                 \\
\multicolumn{1}{l}{Preparatoria Tlajomulco matutino} & \multicolumn{1}{l}{Preparatoria 19 vespertino}     \\
\multicolumn{1}{l}{Bachillerato 9}                  & \multicolumn{1}{l}{Bachillerato 5}                 \\
\multicolumn{1}{l}{Preparatoria 13 matutino}        & \multicolumn{1}{l}{Preparatoria 20 matutino}        \\
\multicolumn{1}{l}{Bachillerato 13}                 & \multicolumn{1}{l}{Bachillerato 11}                 \\
\multicolumn{1}{l}{Preparatoria 4 matutino}         & \multicolumn{1}{l}{Preparatoria 4 vespertino}       \\
\multicolumn{1}{l}{Bachillerato 18}                 & \multicolumn{1}{l}{Bachillerato 19}                 \\
\multicolumn{1}{l}{Preparatoria Tlajomulco vespertino} & \multicolumn{1}{l}{Preparatoria 18 vespertino}      \\
\multicolumn{1}{l}{Bachillerato 1}                  & \multicolumn{1}{l}{Bachillerato 25}                 \\
\multicolumn{1}{l}{Preparatoria 19 matutino}        & \multicolumn{1}{l}{Bachillerato 8}                  \\
\multicolumn{1}{l}{Preparatoria Jalisco matutino}   & \multicolumn{1}{l}{Preparatoria 2 matutino}          \\
\multicolumn{1}{l}{Bachillerato 14} &
\multicolumn{1}{l}{} \\           
\hline
 \insertTableNotes
    \end{longtable}
\end{ThreePartTable}

\setlength{\arrayrulewidth}{0.1mm}
\setlength{\tabcolsep}{12pt}
\renewcommand{\arraystretch}{2.0}
\begin{longtable}{p{2cm}p{2cm}p{2cm}}  
\caption{Proportion of students selected for treatment per school}
\label{tab:school-prop} \\
\hline
 \multicolumn{1}{l}{Order} & \multicolumn{1}{l}{School} & \multicolumn{1}{l}{Proportion} \\ 
\hline 
\multicolumn{1}{l}{1} & \multicolumn{1}{l}{20 VESPERTINO} & \multicolumn{1}{l}{0.195652173913043} \\ 
\multicolumn{1}{l}{2} & \multicolumn{1}{l}{B20} & \multicolumn{1}{l}{0.739130434782609} \\ 
\multicolumn{1}{l}{3} & \multicolumn{1}{l}{5 VESPERTINO} & \multicolumn{1}{l}{0.217391304347826} \\ 
\multicolumn{1}{l}{4} & \multicolumn{1}{l}{17 MATUTINO} & \multicolumn{1}{l}{0.717391304347826} \\ 
\multicolumn{1}{l}{5} & \multicolumn{1}{l}{18 MATUTINO} & \multicolumn{1}{l}{0.0652173913043478} \\ 
\multicolumn{1}{l}{6} & \multicolumn{1}{l}{15 VESPERTINO} & \multicolumn{1}{l}{0.565217391304348} \\ 
\multicolumn{1}{l}{7} & \multicolumn{1}{l}{9 MATUTINO} & \multicolumn{1}{l}{0.58695652173913} \\ 
\multicolumn{1}{l}{8} & \multicolumn{1}{l}{B2} & \multicolumn{1}{l}{0.652173913043478} \\ 
\multicolumn{1}{l}{9} & \multicolumn{1}{l}{B23} & \multicolumn{1}{l}{0.521739130434783} \\ 
\multicolumn{1}{l}{10} & \multicolumn{1}{l}{TLAJOMULCO MATUTINO} & \multicolumn{1}{l}{0.239130434782609} \\ 
\multicolumn{1}{l}{11} & \multicolumn{1}{l}{B17} & \multicolumn{1}{l}{0.391304347826087} \\ 
\multicolumn{1}{l}{12} & \multicolumn{1}{l}{B10} & \multicolumn{1}{l}{0.130434782608696} \\ 
\multicolumn{1}{l}{13} & \multicolumn{1}{l}{13 MATUTINO} & \multicolumn{1}{l}{0.0326086956521739} \\ 
\multicolumn{1}{l}{14} & \multicolumn{1}{l}{4 MATUTINO} & \multicolumn{1}{l}{0.782608695652174} \\ 
\multicolumn{1}{l}{15} & \multicolumn{1}{l}{B30} & \multicolumn{1}{l}{0.978260869565217} \\ 
\multicolumn{1}{l}{16} & \multicolumn{1}{l}{TLAJOMULCO VESPERTINO} & \multicolumn{1}{l}{0.326086956521739} \\ 
\multicolumn{1}{l}{17} & \multicolumn{1}{l}{B9} & \multicolumn{1}{l}{0.141304347826087} \\ 
\multicolumn{1}{l}{18} & \multicolumn{1}{l}{B13} & \multicolumn{1}{l}{0.478260869565217} \\ 
\multicolumn{1}{l}{19} & \multicolumn{1}{l}{B18} & \multicolumn{1}{l}{0.173913043478261} \\ 
\multicolumn{1}{l}{20} & \multicolumn{1}{l}{B1} & \multicolumn{1}{l}{0.891304347826087} \\ 
\multicolumn{1}{l}{21} & \multicolumn{1}{l}{19 MATUTINO} & \multicolumn{1}{l}{0.260869565217391} \\ 
\multicolumn{1}{l}{22} & \multicolumn{1}{l}{JALISCO MATUTINO} & \multicolumn{1}{l}{0.380434782608696} \\ 
\multicolumn{1}{l}{23} & \multicolumn{1}{l}{B14} & \multicolumn{1}{l}{0.695652173913043} \\ 
\multicolumn{1}{l}{24} & \multicolumn{1}{l}{13 VESPERTINO} & \multicolumn{1}{l}{0} \\ 
\multicolumn{1}{l}{25} & \multicolumn{1}{l}{B4} & \multicolumn{1}{l}{0} \\ 
\multicolumn{1}{l}{26} & \multicolumn{1}{l}{5 MATUTINO} & \multicolumn{1}{l}{0} \\ 
\multicolumn{1}{l}{27} & \multicolumn{1}{l}{17 VESPERTINO} & \multicolumn{1}{l}{0} \\ 
\multicolumn{1}{l}{28} & \multicolumn{1}{l}{JALISCO VESPERTINO} & \multicolumn{1}{l}{0} \\ 
\multicolumn{1}{l}{29} & \multicolumn{1}{l}{15 MATUTINO} & \multicolumn{1}{l}{0} \\ 
\multicolumn{1}{l}{30} & \multicolumn{1}{l}{9 VESPERTINO} & \multicolumn{1}{l}{0} \\ 
\multicolumn{1}{l}{31} & \multicolumn{1}{l}{B16} & \multicolumn{1}{l}{0} \\ 
\multicolumn{1}{l}{32} & \multicolumn{1}{l}{B22} & \multicolumn{1}{l}{0} \\ 
\multicolumn{1}{l}{33} & \multicolumn{1}{l}{19 VESPERTINO} & \multicolumn{1}{l}{0} \\ 
\multicolumn{1}{l}{34} & \multicolumn{1}{l}{B27} & \multicolumn{1}{l}{0} \\ 
\multicolumn{1}{l}{35} & \multicolumn{1}{l}{B6} & \multicolumn{1}{l}{0} \\ 
\multicolumn{1}{l}{36} & \multicolumn{1}{l}{20 MATUTINO} & \multicolumn{1}{l}{0} \\ 
\multicolumn{1}{l}{37} & \multicolumn{1}{l}{4 VESPERTINO} & \multicolumn{1}{l}{0} \\ 
\multicolumn{1}{l}{38} & \multicolumn{1}{l}{B3} & \multicolumn{1}{l}{0} \\ 
\multicolumn{1}{l}{39} & \multicolumn{1}{l}{18 VESPERTINO} & \multicolumn{1}{l}{0} \\ 
\multicolumn{1}{l}{40} & \multicolumn{1}{l}{B5} & \multicolumn{1}{l}{0} \\ 
\multicolumn{1}{l}{41} & \multicolumn{1}{l}{B11} & \multicolumn{1}{l}{0} \\ 
\multicolumn{1}{l}{42} & \multicolumn{1}{l}{B19} & \multicolumn{1}{l}{0} \\ 
\multicolumn{1}{l}{43} & \multicolumn{1}{l}{B25} & \multicolumn{1}{l}{0} \\ 
\multicolumn{1}{l}{44} & \multicolumn{1}{l}{B8} & \multicolumn{1}{l}{0} \\ 
\multicolumn{1}{l}{45} & \multicolumn{1}{l}{2 MATUTINO} & \multicolumn{1}{l}{0} \\ 
\hline \\
\end{longtable} 

\subsection{Covariate balance and inclusion}
\label{subsection:cov-balance}

\Cref{tab:cov-balance} includes all 16 covariates considered as controls for the study. Both the educational preferences score and the depression score are unbalanced between groups; however, two of sixteen is something you could expect by chance. They are included in the regressions with all covariates, but show no strong predicting power in educational aspirations of students. It is important to note that, its balance notwithstanding, mother's education, a binary value that takes on the value 1 if the mother has above secondary education and 0 otherwise, is one of the strongest predictors of educational aspirations, consistent with the literature \citep{attanasio2014education}.

\begin{table}[!htbp] \centering 
  \caption{Covariate balance analysis\\
  \textit{Treated vs. Eligible Control}} 
  \label{tab:cov-balance} 
\scalebox{0.75}{\begin{tabular}{@{\extracolsep{5pt}}lccccccc} 
\\[-1.8ex]\hline 
Covariate & \multicolumn{1}{c}{Difference} & \multicolumn{1}{c}{Method} & \multicolumn{1}{c}{P-value} & \multicolumn{1}{c}{Status}\\ 
\hline

Grades & 0.02691 & Welch two sample t-test & 0.341 & balanced\\

Strength & 0.47538 & Welch two sample t-test & 0.5636 & balanced\\ 

Edu. Aspirations (baseline) & N/A & Wilcoxon rank sum test with continuity correction & 0.8826 & balanced\\ 

Edu. Expectations (baseline) & N/A & Wilcoxon rank sum test with continuity correction & 0.6085 & balanced\\ 

Social class [GCN] & 0.158895 & Welch two sample t-test & 0.1351 & balanced\\

Social class [WVS] & N/A & Wilcoxon rank sum test with continuity correction & 0.6899 & balanced\\

Educational preferences & 0.1 & Wilcoxon rank sum test with continuity correction & 0.01706 & unbalanced\\

Risk aversion & -0.163731 & Welch two sample t-test & 0.1032 & balanced \\ 

Depression & 0.08 & Wilcoxon rank sum test with continuity correction & 0.01208 & unbalanced \\

Self-efficacy & N/A & Wilcoxon rank sum test with continuity correction & 0.5436 & balanced \\

Grit & N/A & Wilcoxon rank sum test with continuity correction & 0.3153 & balanced \\

Gender & N/A & Wilcoxon rank sum test with continuity correction & 0.8594 & balanced \\

Mother's education & N/A & Wilcoxon rank sum test with continuity correction & 0.7209 & balanced \\

Transitivity & -0.0011638 & Welch two sample t-test & 0.7586 & balanced \\

Peer effects(baseline) & -0.009633935 & Welch two sample t-test & 0.7022 & balanced \\

\hline \\[-1.8ex] 
\end{tabular}} 
\end{table}

\subsection{Peer effect identification}
\label{subsection:peer-identification}
In their paper, \citet{bramoulle2009identification} prove that the causal identification of peer effects boils down to a property of the matrix $G$: its rank. They propose that as long as the the identity of the matrix $I$, $G$ and the square of the matrix $G^2$ are linearly independent, then we can break the reflection problem in the network parameter $\beta$. i.e. assume that the term is no longer endogenous. By definition, $I$ is linearly independent. $G^2$ is constructed from $G$, and therefore will exhibit similar vector space properties. Thus, we are interested primarily in the rank of $G$, the friendship matrix of all students in the sample. $rank(G) = 6326$ when the full rank is $6348$, the dimension of the full vector space corresponding to the unique number of students in the matrix. There exist 22 vectors that are a linear combination of another vector. However, in their proposition 1, \citet{bramoulle2009identification} show that matrices $I$, $G$ and $G^2$ are linearly dependent if and only if $\E[Gy|x]$ is perfectly collinear with $x$ and $Gx$, where $y$ and $x$ denote the outcome and covariates of interest. From this statement, we can assume that as long as there exists \textit{some} intransitivity (open triangles), we can expect no perfect collinearity between the network outcomes $Gy$ and $x$ and $Gx$. That is, if the diameter of the friendship network $D(G)\geq3$, the peer effects in $\beta$ are identified. In our matrix $D(G) = 20$, therefore there will always exist at least two students separated by a distance of three. \Cref{tab:school-intransitivity} shows the number of open friendship triangles in $G^s$.

\setlength{\arrayrulewidth}{0.1mm}
\setlength{\tabcolsep}{12pt}
\renewcommand{\arraystretch}{2.0}
\begin{longtable}{p{2cm}p{2cm}p{2cm}}  
\caption{Open triangles in the the school network $G^s$}
\label{tab:school-intransitivity} \\
\hline
\multicolumn{1}{l}{School} & \multicolumn{1}{l}{Intransitivity} \\ 
\hline 
\multicolumn{1}{l}{20 VESPERTINO} & \multicolumn{1}{l}{3} \\ 
\multicolumn{1}{l}{B20} & \multicolumn{1}{l}{11} \\ 
\multicolumn{1}{l}{5 VESPERTINO} & \multicolumn{1}{l}{7} \\ 
\multicolumn{1}{l}{17 MATUTINO} & \multicolumn{1}{l}{0} \\ 
\multicolumn{1}{l}{18 MATUTINO} & \multicolumn{1}{l}{1} \\ 
\multicolumn{1}{l}{15 VESPERTINO} & \multicolumn{1}{l}{15} \\ 
\multicolumn{1}{l}{9 MATUTINO} & \multicolumn{1}{l}{6} \\ 
\multicolumn{1}{l}{B2} & \multicolumn{1}{l}{2} \\ 
\multicolumn{1}{l}{B23} & \multicolumn{1}{l}{1} \\ 
\multicolumn{1}{l}{TLAJOMULCO MATUTINO} & \multicolumn{1}{l}{9} \\ 
\multicolumn{1}{l}{B17} & \multicolumn{1}{l}{23} \\ 
\multicolumn{1}{l}{B10} & \multicolumn{1}{l}{4} \\ 
\multicolumn{1}{l}{13 MATUTINO} & \multicolumn{1}{l}{2} \\ 
\multicolumn{1}{l}{4 MATUTINO} & \multicolumn{1}{l}{1} \\ 
\multicolumn{1}{l}{B30} & \multicolumn{1}{l}{4} \\ 
\multicolumn{1}{l}{TLAJOMULCO VESPERTINO} & \multicolumn{1}{l}{1} \\ 
\multicolumn{1}{l}{B9} & \multicolumn{1}{l}{6} \\ 
\multicolumn{1}{l}{B13} & \multicolumn{1}{l}{6} \\ 
\multicolumn{1}{l}{B18} & \multicolumn{1}{l}{12} \\ 
\multicolumn{1}{l}{B1} & \multicolumn{1}{l}{0} \\ 
\multicolumn{1}{l}{19 MATUTINO} & \multicolumn{1}{l}{2} \\ 
\multicolumn{1}{l}{JALISCO MATUTINO} & \multicolumn{1}{l}{6} \\ 
\multicolumn{1}{l}{B14} & \multicolumn{1}{l}{9} \\ 
\multicolumn{1}{l}{13 VESPERTINO} & \multicolumn{1}{l}{2} \\ 
\multicolumn{1}{l}{B4} & \multicolumn{1}{l}{13} \\ 
\multicolumn{1}{l}{5 MATUTINO} & \multicolumn{1}{l}{6} \\ 
\multicolumn{1}{l}{17 VESPERTINO} & \multicolumn{1}{l}{6} \\ 
\multicolumn{1}{l}{JALISCO VESPERTINO} & \multicolumn{1}{l}{7} \\
\multicolumn{1}{l}{15 MATUTINO} & \multicolumn{1}{l}{9} \\ 
\multicolumn{1}{l}{9 VESPERTINO} & \multicolumn{1}{l}{7} \\ 
\multicolumn{1}{l}{B16} & \multicolumn{1}{l}{12} \\ 
\multicolumn{1}{l}{B22} & \multicolumn{1}{l}{1} \\ 
\multicolumn{1}{l}{19 VESPERTINO} & \multicolumn{1}{l}{2} \\ 
\multicolumn{1}{l}{B27} & \multicolumn{1}{l}{2} \\ 
\multicolumn{1}{l}{B6} & \multicolumn{1}{l}{7} \\ 
\multicolumn{1}{l}{20 MATUTINO} & \multicolumn{1}{l}{12} \\ 
\multicolumn{1}{l}{4 VESPERTINO} & \multicolumn{1}{l}{1} \\ 
\multicolumn{1}{l}{B3} & \multicolumn{1}{l}{1} \\ 
\multicolumn{1}{l}{18 VESPERTINO} & \multicolumn{1}{l}{7} \\ 
\multicolumn{1}{l}{B5} & \multicolumn{1}{l}{3} \\ 
\multicolumn{1}{l}{B11} & \multicolumn{1}{l}{8} \\ 
\multicolumn{1}{l}{B19} & \multicolumn{1}{l}{12} \\ 
\multicolumn{1}{l}{B25} & \multicolumn{1}{l}{6} \\ 
\multicolumn{1}{l}{B8} & \multicolumn{1}{l}{6} \\ 
\multicolumn{1}{l}{2 MATUTINO} & \multicolumn{1}{l}{13} \\ 
\hline \\
\end{longtable} 

\subsection{Mechanisms: logistic general models with dynamic peer effects}
Logistic regression for three moments including a variable that expresses whether student i's friends have, on average, changed their aspirations.
\setlength{\arrayrulewidth}{0.1mm}
\setlength{\tabcolsep}{5pt}
\setlength\LTleft{-0.5cm}
\renewcommand{\arraystretch}{1.3}
\label{subsection:mechanism-reg}
\begin{table}[!htbp] \centering
\caption{Log odds: Logistic regression results for three moments [$z_i$] with dynamic peer effects} 
  \label{tab:mechlogistic_yc} 
\begin{tabular}{@{\extracolsep{5pt}}lccc} 
\\[-1.8ex]\toprule 

 & \multicolumn{3}{c}{\textit{Dependent variable:}} \\ 
\cmidrule(l){2-4} 
\\[-1.8ex] & Educ. aspirations $y^c$ 10 & Educ. aspirations $y^c$ 21 & Educ. aspirations $y^c$ 20 \\ 
\\[-1.8ex] & (1) & (2) & (3)\\ 
\midrule \\[-1.8ex]

Treated    & $0.077$       & $0.128$       & $0.027$     \\
               & $(0.047)$     & $(0.072)$     & $(0.047)$   \\
PeerEffects(avg)          & $0.247^{***}$ & $0.362^{***}$ & $0.109^{*}$ \\
               & $(0.053)$     & $(0.084)$     & $(0.053)$   \\
PropTreatFr         & $0.068$       & $-0.058$      & $0.026$     \\
               & $(0.062)$     & $(0.097)$     & $(0.046)$   \\
change\_t10    & $0.416^{***}$ &               &             \\
               & $(0.041)$     &               &             \\
change\_t21    &               & $0.222^{**}$  &             \\
               &               & $(0.080)$     &             \\
change\_t20    &               &               & $0.077$     \\
               &               &               & $(0.069)$   \\
\midrule
Num. obs.      & $515$         & $216$         & $405$       \\
Log Likelihood & $-312.207$    & $-133.901$    & $-257.603$  \\
Deviance       & $624.414$     & $267.801$     & $515.206$   \\
AIC            & $634.414$     & $277.801$     & $525.206$   \\
BIC            & $655.634$     & $294.678$     & $545.225$   \\
\bottomrule
\multicolumn{4}{l}{\scriptsize{Note: $^{***}p<0.001$; $^{**}p<0.01$; $^{*}p<0.05$}}\\
\multicolumn{4}{l}{\scriptsize{Reported standard errors are HC1}}
\end{tabular} 
\end{table}

\subsection{Robustness checks: covariate-controlled logistic models}
\label{subsection:cov-regress}
Covariate-controlled regression tables are in order of appearance throughout the paper.

\setlength{\arrayrulewidth}{0.1mm}
\setlength{\tabcolsep}{5pt}
\setlength\LTleft{-0.5cm}
\renewcommand{\arraystretch}{1.3}
\begin{longtable}{lcccccc}
  \caption{Log-odds: Logistic regression results for three moments [$z_i$]} 
  \label{tab:AME:logistic_yc_covariates} 
\\[-1.8ex]\hline 

 & \multicolumn{3}{c}{\textit{Dependent variable:}} \\ 
\cline{2-4} 
\\[-1.8ex] & Educ. aspirations $y^c$ 10 & Educ. aspirations $y^c$ 21 & Educ. aspirations $y^c$ 20 \\ 
\\[-1.8ex] & (1) & (2) & (3)\\ 
\hline \\[-1.8ex] 
 
Treated                      & $0.123^{***}$ & $0.041$       & $0.037$       \\
                                 & $(0.037)$     & $(0.029)$     & $(0.024)$     \\
                                 &&&\\
aspE0FactPrepa                   & $0.177$       & $0.302$       & $0.453$       \\
                                 & $(0.271)$     & $(0.256)$     & $(0.245)$     \\
                                 &&&\\
aspE0FactTecnico                 & $0.477^{***}$ & $0.579^{***}$ & $0.055$       \\
                                 & $(0.081)$     & $(0.050)$     & $(0.076)$     \\
                                 &&&\\
aspE0FactMtra                    & $0.055$       & $0.065$       & $0.058$       \\
                                 & $(0.067)$     & $(0.068)$     & $(0.061)$     \\
                                 &&&\\
aspE0FactDocto                   & $-0.192$      & $-0.083$      & $-0.027$      \\
                                 & $(0.102)$     & $(0.100)$     & $(0.084)$     \\
                                 &&&\\
Class\_SocialLow income          & $0.673$       & $-0.089$      & $-0.091$      \\
                                 & $(1.368)$     & $(0.248)$     & $(0.164)$     \\
                                 &&&\\
Class\_SocialUpper-middle income & $0.997^{*}$   & $-0.025$      & $-0.078$      \\
                                 & $(0.489)$     & $(0.274)$     & $(0.214)$     \\
                                 &&&\\
Class\_SocialLower-middle income & $0.995$       & $-0.019$      & $-0.069$      \\
                                 & $(0.709)$     & $(0.276)$     & $(0.224)$     \\
                                 &&&\\
Grades                         & $-0.004$      & $-0.024$      & $-0.002$      \\
& $(0.029)$     & $(0.029)$     & $(0.024)$     \\
&&&\\
EduAvg                           & $-0.006$      & $-0.003$      & $-0.025^{*}$  \\
                                 & $(0.016)$     & $(0.015)$     & $(0.012)$     \\
                                 &&&\\
Risk.Pref                     & $-0.008$      & $-0.010$      & $0.001$       \\
                                 & $(0.008)$     & $(0.007)$     & $(0.006)$     \\
                                 &&&\\
DeprAvg                          & $-0.003$      & $0.007$       & $-0.009$      \\
                                 & $(0.019)$     & $(0.018)$     & $(0.015)$     \\
                                 &&&\\
Self.Eff.avg                            & $0.041$       & $-0.028$      & $0.039$       \\
                                 & $(0.034)$     & $(0.033)$     & $(0.028)$     \\
                                 &&&\\
GRITavg                          & $-0.052$      & $-0.035$      & $0.036$       \\
                                 & $(0.030)$     & $(0.028)$     & $(0.024)$     \\
                                 &&&\\
PeerEffects(avg)                            & $-0.168^{*}$  & $-0.079$      & $-0.050$      \\
                                 & $(0.085)$     & $(0.078)$     & $(0.065)$     \\
                                 &&&\\
PropTreatFr                           & $0.028$       & $0.162^{***}$ & $0.119^{***}$ \\
                                 & $(0.022)$     & $(0.023)$     & $(0.019)$     \\
                                 &&&\\
transitivity                            & $0.098$       & $-0.151$      & $0.161$       \\
                                 & $(0.201)$     & $(0.197)$     & $(0.163)$     \\
                                 &&&\\
Gender.M(fem)                    & $-0.010$      & $-0.011$      & $0.010$       \\
                                 & $(0.031)$     & $(0.030)$     & $(0.025)$     \\
                                 &&&\\
MotherEdu                              & $-0.126^{*}$  & $-0.145^{**}$ & $-0.097^{*}$  \\
                                 & $(0.050)$     & $(0.048)$     & $(0.042)$     \\
                                 &&&\\
                                  &&&\\
State(J)                              & $0.022$  & $0.031$ & $0.076^{**}$  \\
                                 & $(0.030)$     & $(0.024)$     & $(0.028)$     \\
                                 &&&\\
\hline
&&&\\
Num. obs.                        & $1242$        & $1242$        & $1242$        \\
Log Likelihood                   & $-746.793$    & $-709.053$    & $-655.551$    \\
Deviance                         & $1493.586$    & $1418.107$    & $1311.103$    \\
AIC                              & $1539.586$    & $1464.107$    & $1357.103$    \\
BIC                              & $1657.449$    & $1581.970$    & $1474.966$    \\
\hline
\multicolumn{1}{l}{\scriptsize{Note: $^{***}p<0.001$; $^{**}p<0.01$; $^{*}p<0.05$}} 
\end{longtable}

\setlength{\arrayrulewidth}{0.1mm}
\setlength{\tabcolsep}{5pt}
\setlength\LTleft{-0.5cm}
\renewcommand{\arraystretch}{1.3}
\begin{longtable}{lcccccc}
  \caption{Log-odds: Logistic regression results for three moments [$z_j^s$]} 
  \label{tab:AME:logistic_yc_covariates2} 
\\[-1.8ex]\hline 

 & \multicolumn{3}{c}{\textit{Dependent variable:}} \\ 
\cline{2-4} 
\\[-1.8ex] & Educ. aspirations $y^c$ 10 & Educ. aspirations $y^c$ 21 & Educ. aspirations $y^c$ 20 \\ 
\\[-1.8ex] & (1) & (2) & (3)\\ 
\hline \\[-1.8ex] 
 
treatedSchool                & $0.041^{**}$       & $0.048^{***}$   & $0.025^{*}$       \\
                                 & $(0.014)$     & $(0.014)$     & $(0.012)$     \\
                                 &&&\\
aspE0FactPrepa                   & $0.278^{*}$   & $0.115$       & $0.033$       \\
                                 & $(0.137)$     & $(0.139)$     & $(0.114)$     \\
                                 &&&\\
aspE0FactTecnico                 & $0.495^{***}$ & $0.542^{***}$ & $0.043$       \\
                                 & $(0.039)$     & $(0.034)$     & $(0.040)$     \\
                                 &&&\\
aspE0FactMtra                    & $0.156^{***}$ & $0.141^{***}$ & $0.052$       \\
                                 & $(0.036)$     & $(0.037)$     & $(0.031)$     \\
                                 &&&\\
aspE0FactDocto                   & $-0.124^{**}$ & $-0.070$      & $-0.068$      \\
                                 & $(0.043)$     & $(0.044)$     & $(0.038)$     \\
                                 &&&\\
Class\_SocialLow income          & $-0.071$      & $-0.174^{**}$ & $-0.020$      \\
                                 & $(0.106)$     & $(0.066)$     & $(0.098)$     \\
                                 &&&\\
Class\_SocialUpper-middle income & $0.054$       & $-0.190^{*}$  & $0.018$       \\
                                 & $(0.112)$     & $(0.094)$     & $(0.096)$     \\
                                 &&&\\
Class\_SocialLower-middle income & $0.070$       & $-0.217^{*}$  & $0.008$       \\
                                 & $(0.108)$     & $(0.101)$     & $(0.094)$     \\
                                 &&&\\
Grades                         & $-0.032^{*}$  & $-0.032^{*}$  & $0.004$       \\
                                 & $(0.014)$     & $(0.015)$     & $(0.012)$     \\
                                 &&&\\
EduAvg                           & $0.006$       & $-0.001$      & $0.001$       \\
                                 & $(0.008)$     & $(0.008)$     & $(0.007)$     \\
                                 &&&\\
Risk.Pref                     & $-0.010^{**}$ & $-0.000$      & $-0.000$      \\
                                 & $(0.004)$     & $(0.004)$     & $(0.003)$     \\
                                 &&&\\
DeprAvg                          & $0.003$       & $-0.010$      & $-0.004$      \\
                                 & $(0.010)$     & $(0.010)$     & $(0.008)$     \\
                                 &&&\\
Self.Eff.Avg                            & $0.030$       & $-0.015$      & $0.025$       \\
                                 & $(0.016)$     & $(0.017)$     & $(0.014)$     \\
                                 &&&\\
GRITavg                          & $-0.040^{**}$ & $-0.034^{*}$  & $-0.027^{*}$  \\
                                 & $(0.015)$     & $(0.015)$     & $(0.012)$     \\
PeerEffect(avg)                            & $-0.084^{**}$  & $-0.089^{**}$  & $-0.060^{*}$  \\
                                 & $(0.032)$     & $(0.032)$     & $(0.027)$     \\
                                 &&&\\
PropTreatFr                           & $0.069^{***}$   & $0.025^{*}$ & $0.049^{***}$ \\
                                 & $(0.011)$     & $(0.010)$     & $(0.009)$     \\
                                 &&&\\
transitivity                            & $0.015$       & $-0.020$      & $-0.024$      \\
                                 & $(0.068)$     & $(0.070)$     & $(0.058)$     \\
                                 &&&\\
Gender.M(fem)                    & $0.004$       & $0.003$       & $-0.000$      \\
                                 & $(0.017)$     & $(0.017)$     & $(0.014)$     \\
                                 &&&\\
MotherEdu                              & $-0.051$      & $-0.067^{**}$      & $-0.032$      \\
                                 & $(0.022)$     & $(0.022)$     & $(0.018)$     \\
                                 &&&\\
State(J)                              & $-0.033^{*}$      & $0.023$      & $-0.018$      \\
                                 & $(0.014)$     & $(0.014)$     & $(0.012)$     \\
                                 &&&\\
\hline
&&&\\
Num. obs.                        & $3734$        & $3734$        & $3734$        \\

Log Likelihood                   & $-2088.253$   & $-2103.778$   & $-1888.645$   \\

Deviance                         & $4176.505$    & $4207.556$    & $3777.289$    \\

AIC                              & $4222.505$    & $4253.556$    & $3823.289$    \\

BIC                              & $4365.686$    & $4396.737$    & $3966.470$    \\
\hline
\multicolumn{4}{l}{\scriptsize{Note: $^{***}p<0.001$; $^{**}p<0.01$; $^{*}p<0.05$}} 
\end{longtable}

\setlength{\arrayrulewidth}{0.1mm}
\setlength{\tabcolsep}{5pt}
\setlength\LTleft{-1cm}
\renewcommand{\arraystretch}{1.3}
\begin{longtable}{lcccccc}
\caption{Log-odds: Multinomial model for three moments [$z_i$]} 
\label{tab:multinomial1cov}\\ \hline

\multicolumn{5}{c}{\textit{Dependent variable:}} \\ 
\cline{2-7} 
\\ & -1 & 1 & -1 & 1 & -1 & 1 \\ 
\\ & $y_i^{dc}$(base 10) & $y_i^{dc}$(base 10) & $y_i^{dc}$(long 20) & $y_i^{dc}$(long 20) & $y_i^{dc}$(end 21) & $y_i^{dc}$(end 21)\\ 
\hline \\

\multicolumn{1}{l}{Treated} & \multicolumn{1}{l}{0.573$^{***}$} & \multicolumn{1}{l}{0.263} & \multicolumn{1}{l}{0.263$^{*}$} & \multicolumn{1}{l}{0.319} & \multicolumn{1}{l}{0.093} & \multicolumn{1}{l}{0.427$^{**}$} \\ 
  & \multicolumn{1}{l}{(0.144)} & \multicolumn{1}{l}{(0.245)} & \multicolumn{1}{l}{(0.145)} & \multicolumn{1}{l}{(0.264)} & \multicolumn{1}{l}{(0.182)} & \multicolumn{1}{l}{(0.181)} \\ 
  & & & & & & \\ 
 \multicolumn{1}{l}{aspE0Fact(Highschool)} & \multicolumn{1}{l}{0.0001$^{***}$} & \multicolumn{1}{l}{2.267$^{*}$} & \multicolumn{1}{l}{0.0001$^{***}$} & \multicolumn{1}{l}{7.810$^{***}$} & \multicolumn{1}{l}{9.841$^{***}$} & \multicolumn{1}{l}{5.935$^{***}$} \\ 
  & \multicolumn{1}{l}{(0.00001)} & \multicolumn{1}{l}{(1.218)} & \multicolumn{1}{l}{(0.00002)} & \multicolumn{1}{l}{(1.298)} & \multicolumn{1}{l}{(1.428)} & \multicolumn{1}{l}{(1.397)} \\ 
  & & & & & & \\ 
 \multicolumn{1}{l}{aspE0Fact(VocTrain)} & \multicolumn{1}{l}{2.182$^{*}$} & \multicolumn{1}{l}{5.730$^{***}$} & \multicolumn{1}{l}{7.261$^{***}$} & \multicolumn{1}{l}{13.626$^{***}$} & \multicolumn{1}{l}{0.904} & \multicolumn{1}{l}{1.796$^{***}$} \\ 
  & \multicolumn{1}{l}{(1.306)} & \multicolumn{1}{l}{(0.415)} & \multicolumn{1}{l}{(1.470)} & \multicolumn{1}{l}{(0.466)} & \multicolumn{1}{l}{(0.558)} & \multicolumn{1}{l}{(0.463)} \\ 
  & & & & & & \\ 
 \multicolumn{1}{l}{aspE0Fact(Master)} & \multicolumn{1}{l}{13.159$^{***}$} & \multicolumn{1}{l}{0.826$^{**}$} & \multicolumn{1}{l}{18.339$^{***}$} & \multicolumn{1}{l}{0.942$^{**}$} & \multicolumn{1}{l}{1.399$^{***}$} & \multicolumn{1}{l}{1.349$^{***}$} \\ 
  & \multicolumn{1}{l}{(0.765)} & \multicolumn{1}{l}{(0.325)} & \multicolumn{1}{l}{(1.038)} & \multicolumn{1}{l}{(0.366)} & \multicolumn{1}{l}{(0.420)} & \multicolumn{1}{l}{(0.404)} \\ 
  & & & & & & \\ 
 \multicolumn{1}{l}{aspE0Fact(PhD)} & \multicolumn{1}{l}{11.187$^{***}$} & \multicolumn{1}{l}{0.000} & \multicolumn{1}{l}{19.398$^{***}$} & \multicolumn{1}{l}{0.000} & \multicolumn{1}{l}{0.766} & \multicolumn{1}{l}{0.978 }\\ 
  & \multicolumn{1}{l}{(0.843)} & \multicolumn{1}{l}{(0.00000)} & \multicolumn{1}{l}{(1.100)} & \multicolumn{1}{l}{(0.000)} & \multicolumn{1}{l}{(0.607)} & \multicolumn{1}{l}{(0.603)} \\ 
  & & & & & & \\ 
 \multicolumn{1}{l}{Clase\_SocialLow income} & \multicolumn{1}{l}{14.923$^{***}$} & \multicolumn{1}{l}{13.514$^{***}$} & \multicolumn{1}{l}{$-$1.415} & \multicolumn{1}{l}{16.662$^{***}$} & \multicolumn{1}{l}{$-$1.435} & \multicolumn{1}{l}{10.594$^{***}$} \\ 
  & \multicolumn{1}{l}{(0.651)} & \multicolumn{1}{l}{(0.865)} & \multicolumn{1}{l}{(1.539)} & \multicolumn{1}{l}{(0.876)} & \multicolumn{1}{l}{(1.459)} & \multicolumn{1}{l}{(0.613)} \\ 
  & & & & & & \\ 
 \multicolumn{1}{l}{Clase\_SocialUpper-middle income} & \multicolumn{1}{l}{16.394$^{***}$} & \multicolumn{1}{l}{14.681$^{***}$} & \multicolumn{1}{l}{$-$0.707} & \multicolumn{1}{l}{15.539$^{***}$} & \multicolumn{1}{l}{$-$1.131} & \multicolumn{1}{l}{10.674$^{***}$} \\ 
  & \multicolumn{1}{l}{(0.412)} & \multicolumn{1}{l}{(0.584)} & \multicolumn{1}{l}{(1.401)} & \multicolumn{1}{l}{(0.634)} & \multicolumn{1}{l}{(1.259)} & \multicolumn{1}{l}{(0.446)} \\ 
  & & & & & & \\ 
 \multicolumn{1}{l}{Clase\_SocialLower-middle income} & \multicolumn{1}{l}{16.293$^{***}$} & \multicolumn{1}{l}{14.916$^{***}$} & \multicolumn{1}{l}{$-$0.874} & \multicolumn{1}{l}{16.219$^{***}$} & \multicolumn{1}{l}{$-$1.206} & \multicolumn{1}{l}{10.856$^{***}$} \\ 
  & \multicolumn{1}{l}{(0.410)} & \multicolumn{1}{l}{(0.575)} & \multicolumn{1}{l}{(1.402)} & \multicolumn{1}{l}{(0.638)} & \multicolumn{1}{l}{(1.258)} & \multicolumn{1}{l}{(0.442)} \\ 
  & & & & & & \\ 
  \multicolumn{1}{l}{Grades} & \multicolumn{1}{l}{$-$0.107} & \multicolumn{1}{l}{0.218} & \multicolumn{1}{l}{$-$0.062} & \multicolumn{1}{l}{$-$0.263} & \multicolumn{1}{l}{0.234} & \multicolumn{1}{l}{$-$0.224} \\ 
  & \multicolumn{1}{l}{(0.141)} & \multicolumn{1}{l}{(0.226)} & \multicolumn{1}{l}{(0.146)} & \multicolumn{1}{l}{(0.253)} & \multicolumn{1}{l}{(0.187)} & \multicolumn{1}{l}{(0.170)} \\ 
  & & & & & & \\  
  
 \multicolumn{1}{l}{EduAvg} & \multicolumn{1}{l}{0.913$^{***}$} & \multicolumn{1}{l}{1.195$^{***}$} & \multicolumn{1}{l}{0.989$^{***}$} & \multicolumn{1}{l}{0.915$^{***}$} & \multicolumn{1}{l}{0.931$^{***}$} & \multicolumn{1}{l}{0.817$^{***}$} \\ 
  & \multicolumn{1}{l}{(0.074)} & \multicolumn{1}{l}{(0.136)} & \multicolumn{1}{l}{(0.077)} & \multicolumn{1}{l}{(0.143)} & \multicolumn{1}{l}{(0.094)} & \multicolumn{1}{l}{(0.088)} \\ 
  & & & & & & \\ 
 \multicolumn{1}{l}{Pref\_Risk} & \multicolumn{1}{l}{0.949$^{***}$} & \multicolumn{1}{l}{1.004$^{***}$} & \multicolumn{1}{l}{0.950$^{***}$} & \multicolumn{1}{l}{0.919$^{***}$} & \multicolumn{1}{l}{1.004$^{***}$} & \multicolumn{1}{l}{1.009$^{***}$} \\ 
  & \multicolumn{1}{l}{(0.036)} & \multicolumn{1}{l}{(0.064)} & \multicolumn{1}{l}{(0.037)} & \multicolumn{1}{l}{(0.073)} & \multicolumn{1}{l}{(0.047)} & \multicolumn{1}{l}{(0.046)} \\ 
  & & & & & & \\ 
 \multicolumn{1}{l}{DeprAvg} & \multicolumn{1}{l}{1.007$^{***}$} & \multicolumn{1}{l}{0.899$^{***}$} & \multicolumn{1}{l}{1.050$^{***}$} & \multicolumn{1}{l}{0.824$^{***}$} & \multicolumn{1}{l}{0.888$^{***}$} & \multicolumn{1}{l}{1.014$^{***}$} \\ 
  & \multicolumn{1}{l}{(0.091)} & \multicolumn{1}{l}{(0.156)} & \multicolumn{1}{l}{(0.092)} & \multicolumn{1}{l}{(0.176)} & \multicolumn{1}{l}{(0.113)} & \multicolumn{1}{l}{(0.115)} \\ 
  & & & & & & \\ 
 \multicolumn{1}{l}{Self.Eff.Avg} & \multicolumn{1}{l}{1.501$^{***}$} & \multicolumn{1}{l}{0.745$^{***}$} & \multicolumn{1}{l}{1.111$^{***}$} & \multicolumn{1}{l}{0.374} & \multicolumn{1}{l}{1.137$^{***}$} & \multicolumn{1}{l}{1.363$^{***}$} \\ 
  & \multicolumn{1}{l}{(0.162)} & \multicolumn{1}{l}{(0.265)} & \multicolumn{1}{l}{(0.168)} & \multicolumn{1}{l}{(0.320)} & \multicolumn{1}{l}{(0.208)} & \multicolumn{1}{l}{(0.203)} \\ 
  & & & & & & \\ 
 \multicolumn{1}{l}{GRITavg} & \multicolumn{1}{l}{0.809$^{***}$} & \multicolumn{1}{l}{0.922$^{***}$} & \multicolumn{1}{l}{0.821$^{***}$} & \multicolumn{1}{l}{1.253$^{***}$} & \multicolumn{1}{l}{1.361$^{***}$} & \multicolumn{1}{l}{1.097$^{***}$} \\ 
  & \multicolumn{1}{l}{(0.135)} & \multicolumn{1}{l}{(0.235)} & \multicolumn{1}{l}{(0.140)} & \multicolumn{1}{l}{(0.261)} & \multicolumn{1}{l}{(0.178)} & \multicolumn{1}{l}{(0.176)} \\ 
  & & & & & & \\ 
 \multicolumn{1}{l}{PeerEffects(avg)} & \multicolumn{1}{l}{-0.762$^{*}$} & \multicolumn{1}{l}{-0.081} & \multicolumn{1}{l}{-0.567 } & \multicolumn{1}{l}{0.615} & \multicolumn{1}{l}{-0.470} & \multicolumn{1}{l}{0.006} \\ 
  & \multicolumn{1}{l}{(0.401)} & \multicolumn{1}{l}{(0.547)} & \multicolumn{1}{l}{(0.408)} & \multicolumn{1}{l}{(0.594)} & \multicolumn{1}{l}{(0.487)} & \multicolumn{1}{l}{(0.478)} \\ 
  & & & & & & \\ 
 \multicolumn{1}{l}{PropTreatFr} & \multicolumn{1}{l}{0.285$^{**}$} & \multicolumn{1}{l}{0.648$^{***}$} & \multicolumn{1}{l}{0.063} & \multicolumn{1}{l}{0.359} & \multicolumn{1}{l}{0.237} & \multicolumn{1}{l}{0.085} \\ 
  & \multicolumn{1}{l}{(0.126)} & \multicolumn{1}{l}{(0.243)} & \multicolumn{1}{l}{(0.122)} & \multicolumn{1}{l}{(0.253)} & \multicolumn{1}{l}{(0.159)} & \multicolumn{1}{l}{(0.154)} \\ 
  & & & & & & \\ 
 \multicolumn{1}{l}{transitivity} & \multicolumn{1}{l}{2.817$^{***}$} & \multicolumn{1}{l}{0.165} & \multicolumn{1}{l}{0.520} & \multicolumn{1}{l}{0.571} & \multicolumn{1}{l}{0.543} & \multicolumn{1}{l}{9.610$^{***}$} \\ 
  & \multicolumn{1}{l}{(0.950)} & \multicolumn{1}{l}{(1.711)} & \multicolumn{1}{l}{(0.992)} & \multicolumn{1}{l}{(1.835)} & \multicolumn{1}{l}{(1.273)} & \multicolumn{1}{l}{(1.155)} \\ 
  & & & & & & \\ 
 \multicolumn{1}{l}{GENDER(male)} & \multicolumn{1}{l}{0.002} & \multicolumn{1}{l}{0.025} & \multicolumn{1}{l}{0.307} & \multicolumn{1}{l}{0.003} & \multicolumn{1}{l}{0.399} & \multicolumn{1}{l}{0.005} \\ 
  & \multicolumn{1}{l}{(0.370)} & \multicolumn{1}{l}{(0.544)} & \multicolumn{1}{l}{(0.686)} & \multicolumn{1}{l}{(0.596)} & \multicolumn{1}{l}{(0.648)} & \multicolumn{1}{l}{(0.409)} \\ 
  & & & & & & \\ 
 \multicolumn{1}{l}{GENDER(female)} & \multicolumn{1}{l}{0.002} & \multicolumn{1}{l}{0.022} & \multicolumn{1}{l}{0.277} & \multicolumn{1}{l}{0.003} & \multicolumn{1}{l}{0.427} & \multicolumn{1}{l}{0.006} \\ 
  & \multicolumn{1}{l}{(0.388)} & \multicolumn{1}{l}{(0.576)} & \multicolumn{1}{l}{(0.704)} & \multicolumn{1}{l}{(0.631)} & \multicolumn{1}{l}{(0.676)} & \multicolumn{1}{l}{(0.432)} \\ 
  & & & & & & \\ 
 \multicolumn{1}{l}{MotherEdu} & \multicolumn{1}{l}{0.514$^{**}$} & \multicolumn{1}{l}{0.765$^{**}$} & \multicolumn{1}{l}{0.571$^{**}$} & \multicolumn{1}{l}{0.408} & \multicolumn{1}{l}{0.652$^{**}$} & \multicolumn{1}{l}{0.570$^{**}$} \\ 
  & \multicolumn{1}{l}{(0.218)} & \multicolumn{1}{l}{(0.341)} & \multicolumn{1}{l}{(0.226)} & \multicolumn{1}{l}{(0.367)} & \multicolumn{1}{l}{(0.264)} & \multicolumn{1}{l}{(0.246)} \\ 
  & & & & & & \\ 
  \multicolumn{1}{l}{State(J} & \multicolumn{1}{l}{0.081} & \multicolumn{1}{l}{0.057} & \multicolumn{1}{l}{0.387$^{**}$} & \multicolumn{1}{l}{0.177} & \multicolumn{1}{l}{0.277} & \multicolumn{1}{l}{-0.009} \\ 
  & \multicolumn{1}{l}{(0.148)} & \multicolumn{1}{l}{(0.256)} & \multicolumn{1}{l}{(0.151)} & \multicolumn{1}{l}{(0.278)} & \multicolumn{1}{l}{(0.190)} & \multicolumn{1}{l}{(0.185)} \\ 
  & & & & & & \\ 
 \multicolumn{1}{l}{Constant} & \multicolumn{1}{l}{-12.236$^{***}$} & \multicolumn{1}{l}{-14.662$^{***}$} & \multicolumn{1}{l}{-0.768 } & \multicolumn{1}{l}{-9.777$^{***}$} & \multicolumn{1}{l}{-2.504 } & \multicolumn{1}{l}{-7.905$^{***}$} \\ 
  & \multicolumn{1}{l}{(1.055)} & \multicolumn{1}{l}{(1.649)} & \multicolumn{1}{l}{(1.629)} & \multicolumn{1}{l}{1.730)} & \multicolumn{1}{l}{(1.769)} & \multicolumn{1}{l}{(1.182)} \\ 
  & & & & & & \\ 

\hline \\ 

Akaike Inf. Crit. & 1,712.909  & 1,712.909 & 1,664.595 & 1,664.595 &  1,851.100 & 1,851.100 \\ 

\hline \\ 

\multicolumn{1}{l}{\scriptsize{Note: $^{*}$p$<$0.1; $^{**}$p$<$0.05; $^{***}$p$<$0.01}} \\
 
\end{longtable}

\subsection{Attrition and handling of missing data}
\label{subsection:attrition}

Participation in the experiment was voluntary for students in selected schools. We therefore expected to see some attrition in the follow-up of the intervention, either as a result of chance, e.g. the student was not in school when the enumerator visited the location, or choice, i.e. the student decides to no longer partake in the study. The total number of students who willingly respond the baseline survey and provide peer network information is $6073$. The total number of students who consistently provide outcome and covariate data for $t \in \{0,1,2\}$ is $4109$, which corresponds to an overall attrition rate of $32.33\%$. A non-zero attrition rate will always have the potential for bias \citep{gerber2012field}. In this section, we present an analysis of the statistical properties of missingness. First, \cref{tab:Descriptive-EduAspirationsMISSING} shows the distribution of the outcome variable, educational aspirations, with the full sample and with a subsample of non-attriters. We observe the same monotonic trend for both the full and the subsample, and the same shift in aspirations from baseline to intervention periods, as reported in the main analysis.

\begin{table}[h!]
\centering
  \resizebox{1\textwidth}{!}{\begin{minipage}{\textwidth}
  \begin{center}
\caption{Educational Aspirations: full sample$^*$ and non-attriters subsample$^{**}$}
\label{tab:Descriptive-EduAspirationsMISSING}

\hspace*{-0.5cm}\begin{tabular}{@{}cccccc@{}}

\toprule
\multicolumn{1}{l}{} & High school & Vocational Training & Undergraduate & Masters & \multicolumn{1}{l}{PhD}  \\ 
\bottomrule
\multicolumn{1}{l}{Baseline$^*$}  & \multicolumn{1}{l}{25} & \multicolumn{1}{l}{422} & \multicolumn{1}{l}{554}  & \multicolumn{1}{l}{1005} & 4066 \\ 
\multicolumn{1}{l}{Intervention$^*$} & \multicolumn{1}{l}{21} & \multicolumn{1}{l}{71} & \multicolumn{1}{l}{998} & \multicolumn{1}{l}{1061} & 2930 \\ 
\multicolumn{1}{l}{Follow up$^*$} & \multicolumn{1}{l}{23} & \multicolumn{1}{l}{58}  & \multicolumn{1}{l}{1009} & \multicolumn{1}{l}{1065} & 2541 \\ \midrule \\ 
\midrule \\
\multicolumn{1}{l}{Baseline$^{**}$}  & \multicolumn{1}{l}{15} & \multicolumn{1}{l}{280} & \multicolumn{1}{l}{366}  & \multicolumn{1}{l}{675} & 2773 \\ 
\multicolumn{1}{l}{Intervention$^{**}$} & \multicolumn{1}{l}{17} & \multicolumn{1}{l}{58} & \multicolumn{1}{l}{782} & \multicolumn{1}{l}{867} & 2385 \\ 
\multicolumn{1}{l}{Follow up$^{**}$} & \multicolumn{1}{l}{18} & \multicolumn{1}{l}{47}  & \multicolumn{1}{l}{889} & \multicolumn{1}{l}{950} & 2194 \\
\bottomrule
\end{tabular}
\end{center}
\end{minipage}}
\end{table}

\begin{figure}[h]
\begin{center}
\caption{Predicted probabilities of $r_i$ given covariate row vector $x'_i$}
\label{figure:predprobhist}
\includegraphics[width=0.7\textwidth]{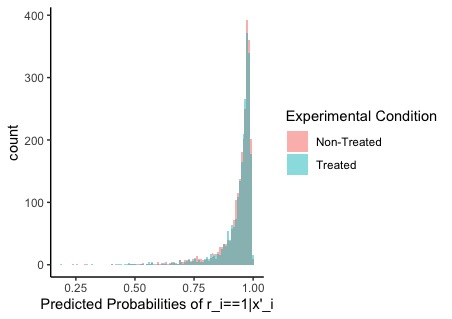}
\end{center}
\end{figure}

\paragraph{}
Further, let $r_i$ be a binary indicator for whether student $i$ reported outcome data at follow-up, such that $pr(r_i=1|z_i)$ is the probability of reporting outcome data given the student's treatment assignment $z_i$, and $pr(r_i=1|x'_i)$ is the probability of reporting outcome conditional on a row vector of pre-treatment coviariates. \Cref{figure:predprobhist} shows histograms of the predicted probabilities of reporting outcome data by treatment assignment, conditional on a row vector of pre-treatment covariates. The binary indicator $r_i$ for \cref{figure:predprobhist} was constructed based on the overall attrition rate, baseline to endline, 32.33\%. The included covariates are baseline educational aspirations and expectations, social class, grades, gender, endogenous peer effect (average aspirations of friend group), mother's education, transitivity, and psychological characteristics (depression, grit, self-efficacy, educational preferences, risk aversion); this is the same pre-treatment covariate row vector that is used for the covariate-controlled regressions. The overlapping distribution of the predicted probabilities of reporting for treated and non-treated students suggests that the data is Missing Independent of Potential Outcomes (MIPO). In this specific case, however, treatment assignment $z_i$ considers treated students as 1, and non-treated students in treated schools and non-treated schools as zero.   

\paragraph{}
As with the main study, we divide the sample into two: the experimental subsample, which includes only students eligible for treatment in treated and non-treated schools, and the subsample of students that are not eligible for treatment in both sets of schools, that served to analyse treatment spillovers. For the experimental subsample,  $pr(r_i=1|z_i) = 0.019$, when $r_i$ denotes reporting behaviour from baseline to endline. With $p = 0.329$, we can safely assume that attrition is uncorrelated with the treatment assignment for treatment-eligible students. For the subsample of non-eligible students, $pr(r_i=1|z_i) = 0.018$ and $p = 0.068$, thus attrition is also uncorrelated with treatment assignment. Reported regression results in the main analysis are based on non-attriters or, as per \citet{gerber2012field} terminology, always-reporters, making the ATE unbiased estimates. 

\end{document}